\pdfoutput=1
\documentclass[manuscript, screen]{acmart}
\AtBeginDocument{%
  }

\setcopyright{acmlicensed}
\copyrightyear{2018}
\acmYear{2018}
\acmDOI{XXXXXXX.XXXXXXX}

\acmJournal{JACM}
\acmVolume{37}
\acmNumber{4}
\acmArticle{111}
\acmMonth{8}

\usepackage{threeparttable}

\usepackage{fancyhdr}

\usepackage{textcomp}
\usepackage{wrapfig,xspace,comment,color}

\usepackage{amsfonts}
\usepackage{algorithmic}
\usepackage[]{algorithm2e}
\usepackage{multirow}
\usepackage{multicol}
\usepackage{listings}
\usepackage{enumitem}
\setitemize{noitemsep,topsep=0pt,parsep=0pt,partopsep=0pt}
\usepackage{xcolor}
\usepackage{url}
\usepackage{hyperref}
\usepackage{mathtools}
\usepackage{amsthm}
\theoremstyle{plain}
\usepackage{float}

\newcommand{\tb}[1]{\textcolor{black}{#1}}

\newcommand{\fpguard}{{\sc FpGuard}\xspace}

\newcommand{\seesaw}{{\sffamily\scshape Seesaw}\xspace}

\newcommand{\satire}{{\sffamily\scshape Satire}\xspace}

\newcommand{\precisa}{{\sffamily\scshape Precisa}\xspace}
\newcommand{\fptaylor}{{\sffamily\scshape Fptaylor}\xspace}
\newcommand{\fluctuat}{{\sffamily\scshape Fluctuat}\xspace}
\newcommand{\fptuner}{{\sffamily\scshape Fptuner}\xspace}
\newcommand{\precimonious}{{\sffamily\scshape Precimonious}\xspace}

\begin{document}

\title{\satire: Computing
Rigorous Bounds for
Floating-Point
Rounding Error in
Mixed-Precision
Loop-Free Programs}

\author{Tanmay Tirpankar}
\email{tanmay.tirpankar@utah.edu}
\orcid{1234-5678-9012}
\affiliation{%
  \institution{University of Utah}
  \city{Salt Lake City}
  \state{Utah}
  \country{USA}
}

\author{Arnab Das}
\affiliation{%
  \institution{University of Utah}
  \city{Salt Lake City}
  \state{Utah}
  \country{USA}
}

\author{Ganesh Gopalakrishnan}
\affiliation{%
  \institution{University of Utah}
  \city{Salt Lake City}
  \country{USA}}
\email{ganesh@cs.utah.edu}

\begin{abstract}
  Techniques
that rigorously
bound 
the overall rounding error exhibited by
a numerical
program are of 
significant interest
for communities developing numerical
software.
However,  
there are very few 
available tools today that
can  be 
used to rigorously bound
errors in
programs 
that
employ conditional statements
(a basic need)
as well as mixed-precision
arithmetic (a direction of significant future interest) employing global optimization in error analysis.
In this paper, we present a new tool\footnotemark that fills this void
while also employing
an abstraction-guided optimization
approach to allow designers to 
trade error-bound tightness for
gains in analysis time---useful when searching for design alternatives.
We first present the basic rigorous analysis framework
of \satire and then show how to extend 
it to incorporate
abstractions, conditionals, and mixed-precision arithmetic.
We begin by describing \satire's 
design and its performance
on a collection of benchmark 
examples.
We then describe these aspects
of \satire:
(1)~how the error-bound and  tool execution time vary with the abstraction level; 
(2)~the additional
machinery
to handle
conditional
expression
branches,
including
defining the
concepts of 
instability jumps
and instability window widths
and measuring 
these quantities;
and (3)~how the error changes when a mix of precision values are used.
To showcase how \satire
can add value  during design,
we start with a 
Conjugate Gradient solver and demonstrate how its 
step size  
and search direction
  are affected
by different precision settings.
\satire is freely available for evaluation,
and can 
be used
during the design of numerical routines to
effect
design tradeoffs guided by rigorous empirical error guarantees.



\footnotetext{\textbf{Extension of Conference Papers.} 
We believe that the {\bf 30\%} extra
material required of a journal version is more than achieved by combining the results from the main \satire-related paper~\cite{2020_Das}
that won the best student paper award in SC 2020,
incorporating results from our Cluster'21 paper~\cite{2021_Das} where handling of branches was introduced, adding the new work on mixed-precision
based on ideas in our former paper~\cite{fptuner}
and  showcasing via a Conjugate Gradient example based on a tool extension to analyze error across mixed-precision.}
\end{abstract}



\maketitle

\section{Introduction}
\label{sec:intro}


Numerical programs are designed
by programmers who follow their
intuitions with
respect to real arithmetic.
Unfortunately, these intuitions cannot easily scale when analyzing even  simple expressions.
Consider for
instance a conditional assignment where the variables are instantiated in single-precision (FP32):
\(
y = \left( (x_1^2 + x_2^2 \leq 10) \, ? \, (0.1 \cdot x_1) \, : \, (0.2 \cdot x_2) \right), \; \)
 \(  x_1, x_2 \in [0.1,5.0]
\) and a few simple questions around it:
(1)~What is $y$'s  maximum error (an empirically guaranteed and tight upper-bound), considering both the outcomes of the conditional? 
(2)~Given that the predicate computation incurs floating-point rounding error whose value can only be conservatively estimated, how much of a ``grey zone'' ({\em instability
window width}) do we have (the then/else cases having an overlap due to rounding errors in conditionals).
One way to  answer these questions
is  by
evaluating
this expression
on pairs 
of
FP32 inputs and also on a version of the expression in FP64 precision (``shadow evaluation'') to determine the rounding error (difference between the FP32 and FP64 values of $y$).
There are over 
$2\times 10^{15}$
FP32 input pairs
in the indicated
ranges, making this brute-force testing applicable to only very short programs.
Now if the user wants to 
know the consequences of
 carrying out $x_1^2$ 
 in double-precision
 (FP64) and downcasting  the result to single-precision
 (FP32) before addition with $x_2^2$,
they would need to run the 
program in a shadow precision that exceeds both these precision values (e.g.,  run it in FP128).

\begin{figure}
    \centering
    \includegraphics[width=0.75\linewidth]{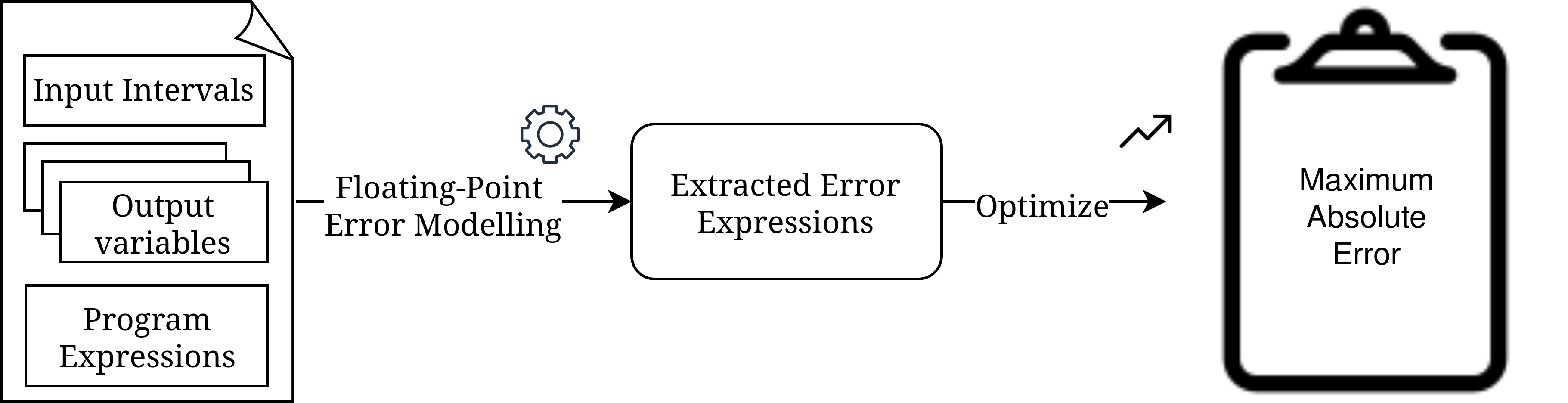}
    \caption{Error Analysis Workflow using Error Expression Optimization}
    \label{fig:intro-workflow}
\end{figure}

Observe Figure~\ref{fig:intro-workflow} which shows the workflow of tools that utilize error expression optimization for error analysis. \satire is one such tool using which, one can obtain conservative estimates at {\em all the points in the domain} without search. For our example, these answers can be obtained almost instantaneously: (1)~The error-bound is 
$5.96\times 10^{-8}$ irrespective of the precision configuration of this short expression (later experiments demonstrate the variety over larger expressions);
(2)~When using
single-precision exclusively,
the width of the instability window is $5.96\times 10^{-6}$,
whereas 
using mixed-precision, 
it is
$1.49\times 10^{-6}$,
and 
under double-precision,
it is 
$1.0\times 10^{-14}$;
and finally (3)~The instability jump 
(magnitude of change)
when switching over 
from the
``then'' case to the ``else'' case is
$0.99$ for both precision
choices.
%
%
In fact,
\satire can 
produce analysis results
of this nature
in about an hour on a consumer grade desktop machine even for expressions with about 4 million operators (scalability depends on the nature of expressions, as will be described).
In contrast, \satire's  well-known predecessor \fptaylor~\cite{fptaylor} can only handle about 200 operators while not handling conditionals.
In our experience, \satire's conservative error bounds are tight-enough to be used 
while exploring design tradeoffs such as precision and the nature of floating-point expressions.

\noindent{\bf Lineage of \satire, and its Key Features:\/}
The work on \satire
began with the efforts reported
in~\cite{2020_Das}\footnotemark
where its scalability as well as error-bound tightness were first published.
\footnotetext{\satire stands for Scalable Abstraction guided Tool for RIgorous Error analysis.}
Ideas in this 2020
version of \satire were
extended in~\cite{2021_Das}
in a tool called
\seesaw which handles
conditional
expressions.
At this stage,
we still did not have
the ability
to handle mixed-precision 
designs.
Also, we had not
showcased our
tools within realistic
design contexts.
This paper combines our prior disparate
tool developments and also makes
significant advances on top of this combination.
Specifically, it 
combines the previous \satire and \seesaw
tools, adds mixed-precision analysis support
and showcases the resulting tool (also
named \satire for continuity)
in a realistic
design context---namely, the design of a simple Conjugate Gradient solver.
Our results demonstrate that
using \satire, one can
study the impact of numerical precision on
step size ($\alpha$)
and search direction
($\beta$), 
thus demonstrating
\satire's ability to help
during realistic design space exploration.\footnotemark
%
%

 %


%

%
%

\noindent{\bf Key Innovations in \satire~\cite{2020_Das} and \seesaw~\cite{2021_Das}:\/}
\satire~\cite{2020_Das}'s key advance over \fptaylor
was 
to give up on second-order error analysis.
First of all, second-order error is of the order of the value of subnormal floating-point numbers
which are less than
$1.175\times 10^{-38}$ for FP32.
While second-order
error analysis may
occasionally be 
needed in analyzing
functions
such as $log$ (e.g.,
see \cite{aiken-math-dot-h}), for the most part
it does not matter, especially considering that our error analysis is conservative.
%
%
During our 2020 work on \satire,
we observed 
a key advantage of not pursuing
second-order analysis as done 
in \fptaylor: 
one does not 
need to
compute output errors {\em on top of
the incoming first-order
errors}
(e.g., see
the error derivations  in~\cite{darulova-phd} which presents a way to sum over all orders of errors).
This observation
allowed us
to carry out error
analysis in a localized
and decoupled manner
by sticking with first-order
error analysis, and employing
a dynamic programming based 
algorithm that sums over first-order errors.
Specifically,
we can conduct a 
backward pass through a computational
graph and accumulate 
reverse-mode derivatives at the visited nodes.
In contrast, 
\fptaylor 
faces an exponentially growing
number of paths over which to sum the error---causing its runtime to increase quickly.
Despite giving up on second-order error, \satire manages to produce tighter error bounds than \fptaylor 
through other 
innovations---specifically
through the use of 
expression canonicalization that actually helps the backend optimizer arrive at far tighter error-bound estimates.
%
These
gains
allow \satire to
scale  as well as provide tight error bounds.
Another key innovation
in \satire is a novel abstraction scheme that enables
a user to
strike a good balance between error-bound tightness and analysis time.

\noindent{\bf Key Innovations, Roadmap:\/} 
We begin with
\S\ref{sec:bg} that provides relevant background
and follow it with
\S\ref{sec:overview} 
that explains \satire's workflow.
We then discuss five of \satire's key innovations:
\begin{itemize}
    \item \textbf{\textit{Path Strength Reduction}} exponentially reduces  the effort for calculating error bounds (\S\ref{sec:psr}). 
    \item \textbf{\textit{Error-Bound Optimization}} reformulates error expressions into a canonical form, enabling global optimization to produce tighter bounds over the input space (\S\ref{sec:canonicalize}). 
    \item \textbf{\textit{Computation Graph Abstraction}} employs a divide-and-conquer strategy, leveraging information-theoretic heuristics to isolate large sub-expressions and summarize their errors (\S\ref{sec:abstractions}).
    
    \item \textit{\textbf{Error Accumulation Over Conditionals}} calculates the worst-case error bounds in presence of conditionals and determines whether the control flow paths diverge when a program is studied under the real number model versus its floating-point and how much the answer differs.
    (\S\ref{sec:error-accumulation}).
    
    \item \textbf{\textit{Mixed-Precision Extension}} 
    calculates  errors across different precision regimes (\S\ref{sec:mixed-precision}).
\end{itemize}

Thereafter, in 
\S\ref{sec:eval}, we
provide our evaluation results based on our earlier work
but using the latest implementation of \satire (rerunning
all experiments).
\S\ref{sec:mix-prec-eval} provides 
our evaluation 
results that 
specifically focus on mixed-precision. 
Then
in \S\ref{sec:conclusion},
we discuss additional
related work, future work and concluding remarks.

\section{Background}
\label{sec:bg}

We now provide background on floating-point arithmetic, Taylor expansion
of floating-point expressions
to compute rounding error, 
reverse-mode symbolic differentiation to propagate the error to the output,
and bounding the error expression using rigorous optimization tools.


\subsection{Floating-point Arithmetic}
 A floating-point number 
   (see~\cite{Goldberg1991, 2010_Muller} for details)
 $\mathbb{F}$ with radix, $\beta$=2 is a binary number system, representing a
subset of real numbers 
using finite precision
bit-strings $(s, e, m)$,
where $s$ (sign-bit)
has values in $\{-1, 1\}$, 
$e$ has the exponent value specific to the number format (FP32, FP64, etc), and $m$ has the mantissa value (when prepended with the implied $1$ for normal numbers and $0$ for subnormals).
The number
represented has real value 
$s \cdot m \cdot \beta^{e}$
in  $\mathbb{F} \subset \mathbb{R} $.
As concrete examples, half-precision (FP16) expresses $e$ in five bits
with values in $[-14,15]$ and 
 $m$ in 10 bits
with values in
$[0, (2-2^{-10})]$.
For single-precision (FP32),
the width of $e$
is $8$ bits and its range is
$[-126,127]$, with $m$ occupying
23 bits, with values in 
$[0, (2-2^{-23})]$.
For double-precision (FP64),
$e$ is 11 bits wide and in the range $[-1022,1023]$ with $m$ being 52 bits wide, with values in 
$[0, (2-2^{-52})]$.
Values of $0 < m < 1.0$ 
and $e$ below the ranges indicated here  (this is the ``all $e$ bits are $0$'' case, since $e$ is stored with a bias, with the bias for FP32 being 127, as an example)
belong to the subnormal range.
In this work, we do not consider subnormals.
Values of $e$ above the range indicated here (this is the ``all bits of $e$ being 1'' case) belong to exception codes (NaN or Not-a-Number and INF or Infinity)---again ignored here.
Finally, if $m=0$ and  all $e$-bits are $0$, we
represent $+0$ or $-0$ 
depending on the sign bit.

For all $x \in \mathbb{R}$,
 $\Tilde{x}$
denotes the element in $\mathbb{F}$ closest to x  with $(x- \Tilde{x})$
being the absolute error and $(x- \Tilde{x})/x$ being the 
relative error.
Every real number $x$ lying in  the range of $\mathbb{F}$ can be approximated by   $\Tilde{x} \in \mathbb{F}$ with a relative error no larger than   $\mathbf{u} = 2^{-(p+1)}$ (called the unit round-off) where p is the number of bits required to represent the mantissa for a given precision (see earlier discussion). Thus, $\forall x \in \mathbb{R},  \Tilde{x} = x(1+\delta)$ where   $|\delta| \leq \mathbf{u}$
(this assumes the {\em round-to-nearest with ties to even} rounding mode).
%
 %
  For each floating-point operator $\diamond \in \{+,-,\times,/\}$, IEEE-standard compatible implementations
  guarantee that the relative error introduced by these operations  will not exceed \textbf{u}~\cite{2008_zuras, 2019_loosemore}. In other words, the result is equal to the exact value $\tilde{x} \diamond \tilde{y}$, plus an error term $\mathcal{E}^{lr} = (\tilde{x} \diamond \tilde{y})\delta$
representing the amount of local relative
round-off error (``$lr$'').
For other rounding modes and operators outside of
the set above, $\mathbf{u}$ is increased---typically doubled for other rounding modes.
If we use $\circ$ to denote rounding, we have

\begingroup
\begin{equation} \label{eq:floating-point-operation}
    \begin{split}
        \circ(\Tilde{x} \diamond \Tilde{y}) = (\Tilde{x} \diamond \Tilde{y})(1+ \delta) = (\Tilde{x} \diamond \Tilde{y}) + (\Tilde{x} \diamond \Tilde{y})\delta, |\delta| \leq \mathbf{u}
    \end{split}  
\end{equation}
\endgroup

    

\subsection{Real Expressions}

Let $\mathscr{E}$
denote a real-valued expression 
whose inputs are a product of bounded intervals, i.e. $x \in \mathbb{I}$ where $\mathbb{I}=[a_1, b_1]\times\cdots\times[a_n, b_n]$ and $a_i \leq x_i \leq b_i$. 
The corresponding floating-point expression $\Tilde{\mathscr{E}}$ contains operators 
that
will be instantiated to
different precision values
(``mixed-precision''); here we assume
uniform precision to keep notations simple.
We will build a syntax-tree and employ a post-order numbering (which tracks dependency order) for operators of all sub-expressions in $\mathscr{E}$. We use $\dot{\mathscr{E}}$ to refer to the operator at the root of the tree of $\mathscr{E}$. Similarly, we use $\mathscr{S}$ to refer to a sub-expression in $\mathscr{E}$ and $\dot{\mathscr{S}}$ to refer to its principal operator.

\subsection{Taylor Expansion of Error}
\label{subsec:taylor_expansion}

Now consider a straight-line program with $n$ floating-point sub-expressions with $\dot{\mathscr{S}}$ ranging over arithmetic operators and elementary functions finally yielding the output $\mathscr{E}_n$ with each line of the form: $\mathscr{S}_i = \dot{\mathscr{S}_i}(x_1, \dotsc, x_m, \mathscr{S}_1, \mathscr{S}_2,\dots, \mathscr{S}_{i-1})$.
%
Let function $\dot{\mathscr{S}_i}$ be approximated
in finite precision by $\tilde{\mathscr{S}_i}$.
Then   $\tilde{{\mathscr{S}_i}}$  
can be 
elaborated
as follows.
We use
$\mathcal{E}^{lr}_{\mathscr{S}_i}$ to represent the round-off error introduced by
$\tilde{\dot{\mathscr{S}_i}}$ where the super-script ``lr'' stands for local-rounding and ``tr'' (used later) stands for total-rounding.
$\tilde{\mathscr{S}_i}$ depends on both the rounded inputs $\tilde{x}_1, \dotsc, \tilde{x}_m$ and earlier intermediate values $\tilde{\mathscr{S}_1}, \dotsc, \tilde{\mathscr{S}}_{i-1}$;
or we can instead think of $\tilde{\mathscr{S}_i}$ as depending on the exact inputs and intermediates $x_1, \dotsc, x_m$ and $\mathscr{S}_1, \dotsc, \mathscr{S}_{i-1}$
{\em and} their rounding errors
$\mathcal{E}^{lr}_{x_1}, \dotsc, \mathcal{E}^{lr}_{x_m}$ and $\mathcal{E}^{lr}_{\mathscr{S}_1}, \dotsc, \mathcal{E}^{lr}_{\mathscr{S}_{i-1}}$:
\begin{align*}
	\tilde{\mathscr{S}_i} 
	&= \tilde{\dot{\mathscr{S}_i}}(\tilde{x}_1, \dotsc, \tilde{x}_m, \tilde{\mathscr{S}_1}, \dotsc, \tilde{\mathscr{S}}_{i-1}) \\
	&= \tilde{\dot{\mathscr{S}_i}}(x_1+\mathcal{E}^{lr}_{x_1},\dots,x_m+\mathcal{E}^{lr}_{x_m}, \mathscr{S}_1 + \mathcal{E}^{lr}_{\mathscr{S}_1}, \dotsc, \mathscr{S}_{i-1} + \mathcal{E}^{lr}_{\mathscr{S}_{i-1}}) \\
    &= \dot{\mathscr{S}_i}(x_1+\mathcal{E}^{lr}_{x_1},\dots,x_m+\mathcal{E}^{lr}_{x_m}, \mathscr{S}_1 + \mathcal{E}^{lr}_{\mathscr{S}_1}, \dots, \mathscr{S}_{i-1} + \mathcal{E}^{lr}_{\mathscr{S}_{i-1}})( 1 + \delta_i);
	\quad |\delta_i| \leq \bf{u}
\end{align*}
Let $\mathcal{E}^{tr}_\mathscr{E}$ be the total error at $\mathscr{E}$. We can now express the error $\mathcal{E}^{tr}_{\mathscr{E}_{n}}$ of the program output as    ``computed minus true,'' {\em i.e.},
\begingroup
\small
\begin{equation*}
\begin{split}
	\mathcal{E}^{tr}_{\mathscr{E}_{n}}
	&= \dot{\mathscr{E}_n}(x_1+\mathcal{E}^{lr}_{x_1},\dots,x_m+\mathcal{E}^{lr}_{x_m},\mathscr{S}_1 + \mathcal{E}^{lr}_{\mathscr{S}_1}, \dots, \mathscr{S}_{i-1} + \mathcal{E}^{lr}_{\mathscr{S}_{n-1}})( 1 + \delta_n) - \mathscr{E}_n;\quad |\delta_n| \leq \bf{u}
\end{split}
\end{equation*}
\endgroup

The Taylor expansion of the final error expression provides an efficient way to bound this error.
Trivially, $\mathcal{E}^{tr}_{\mathscr{E}_n}$ is zero
if the error values $\mathcal{E}^{lr}_{\mathscr{S}_i}$, $\mathcal{E}^{lr}_{x_i}$, and $\delta_n$ are all equal to zero.
This justifies a Taylor expansion about zero in those error values.
So finally, we have this form for the total error, which 
includes the error introduced at the final expression
$\mathscr{E}_n$ 
and
 the first order errors
 at all earlier stages 
$\mathscr{S}_j$
and the 
first order errors at 
inputs $x_j$ (the second order 
error components $O({\bf u}^2)$ are ignored in our analysis).
\begingroup
\small
\begin{equation}
\label{eq:taylor-expansion}
   \mathcal{E}^{tr}_{\mathscr{E}_{n}} = \mathscr{E}_n\delta_n
            + \sum_{j=1}^{n-1} \dfrac{\partial \mathscr{E}_n}{\partial \mathscr{S}_j}\mathcal{E}^{lr}_{\mathscr{S}_j}
            + \sum_{j=1}^{m} \dfrac{\partial \mathscr{E}_n}{\partial x_j}\mathcal{E}^{lr}_{x_j}
            + O({\bf u}^2)
\end{equation}
\endgroup

\subsection{Error Bounding}
\label{subsec:error_bounding}

\satire's goal is to bound $\mathcal{E}^{tr}_{\mathscr{E}_{n}}$, the {\em total error} of the program.
From Equation$~\eqref{eq:taylor-expansion}$, we can observe that $\mathcal{E}^{tr}_{\mathscr{E}_{n}}$ is composed of the {\em local error} generated by the application of $\dot{\mathscr{E}_n}$, which we write ${\mathcal E}^{lr}_{\mathscr{E}_{n}}$, and the propagation of the incoming errors $\mathcal{E}^{lr}_{x_j}$ and $\mathcal{E}^{lr}_{\mathscr{S}_j}$~\cite{rosa,higham}.
The propagation strength is determined by the partial derivative terms.
Given intervals $x \in \mathbb{I}$, the {\bf error-bound} on $\mathcal{E}^{tr}_{\mathscr{E}_{n}}$,
which we write $E^{tr}(\mathscr{E}_n)$,
can be computed as follows:

\begingroup
\small
\begin{equation}
\begin{split}
E^{tr}(\mathscr{E}_{n}) &\leq \max_{{\bf x}\in I({\bf x})}(|\mathcal{E}^{tr}_{\mathscr{E}_{n}}|)  
	 \leq \max_{{\bf x}\in I({\bf x})}
         \bigg (
|\mathcal{E}^{lr}_{\mathscr{E}_n}| + |\sum_{j=1}^{n-1} \dfrac{\partial \mathscr{E}_n}{\partial \mathscr{S}_j}
      \mathcal{E}^{lr}_{\mathscr{S}_j}| + |\sum_{j=1}^{m} \dfrac{\partial \mathscr{E}_n}{\partial x_j}\mathcal{E}^{lr}_{x_j} | 
					\bigg ) + O({\bf u}^2) \\
\end{split}
\label{eq:total-final-form}
\end{equation}
\endgroup

This equation says that the rounding error 
can be upper-bounded by invoking an optimizer
that finds the $max$ values within the indicated
intervals.
The summations in the error function are accumulated using 
\satire's dynamic programming procedure.


\subsection{Reverse-Mode Symbolic Differentiation}

\begin{figure*}[h]
    \begin{minipage}{0.48\textwidth}
    \begin{center}
    \includegraphics[width=0.5\linewidth, alt={Illustration of Reverse Mode AD}]{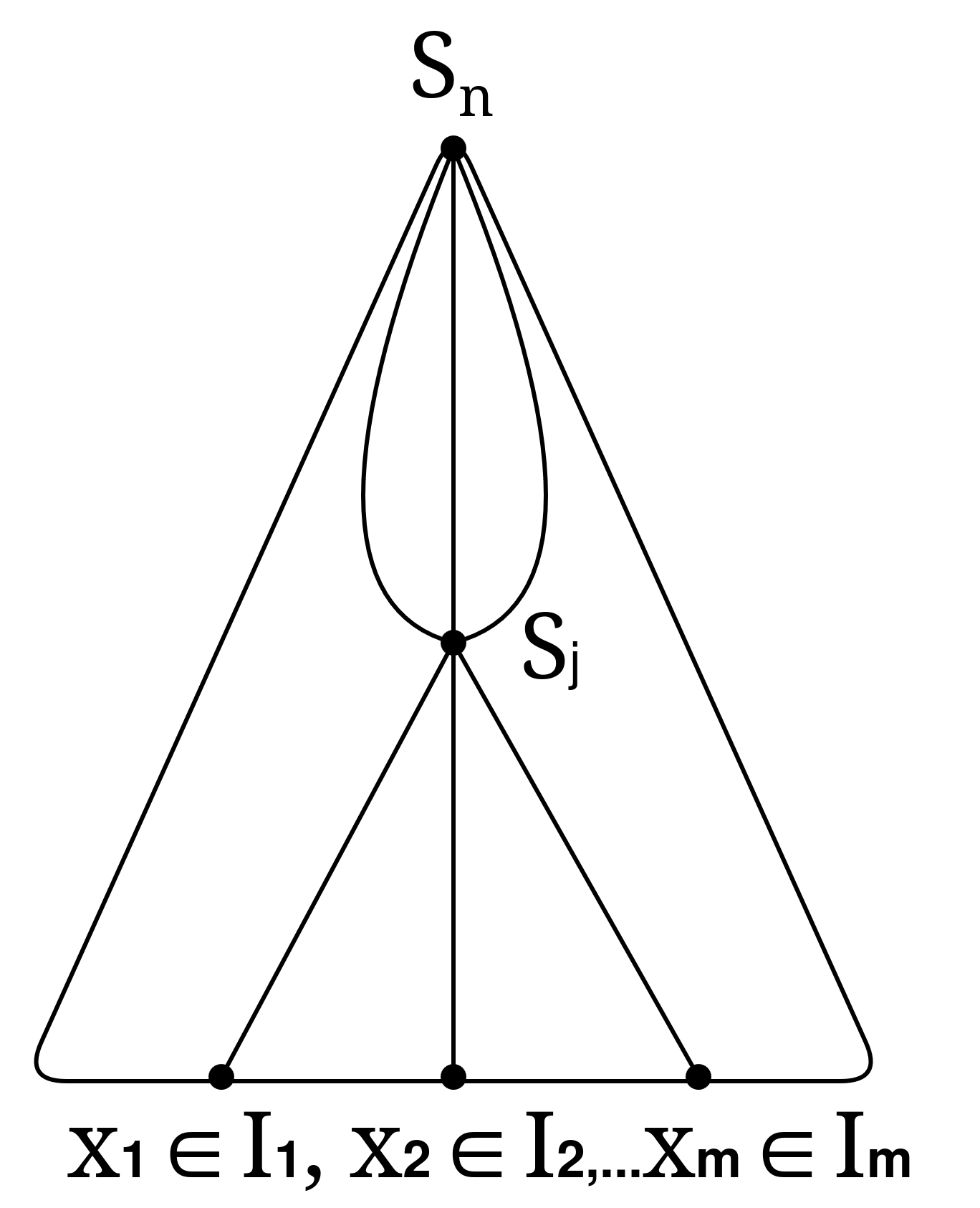}
    \end{center}
  \caption{\label{fig:reverse-mode-ad} Reverse Mode AD}
  \end{minipage}
  \hfill
  \begin{minipage}{0.48\textwidth}
  \begin{center}
   \includegraphics[width=1.0\linewidth]{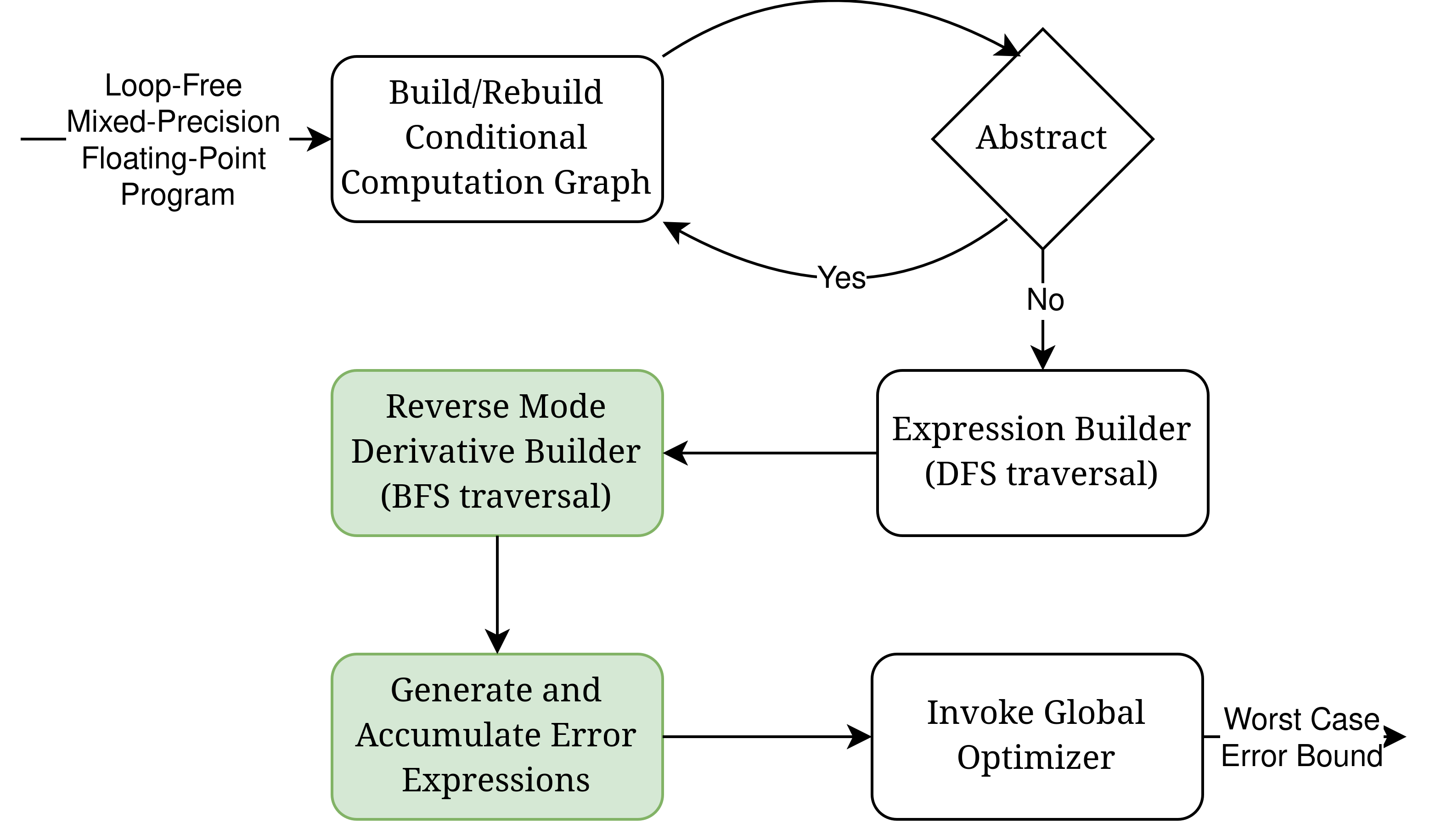}
   \end{center}
   \caption{Overview of \satire} \label{fig:overview}
  \end{minipage}
\end{figure*}

The floating-point expressions comprising  the straightline programs we discussed thus far
are modeled as computational graphs, with the partial derivatives computed through reverse-mode symbolic differentiation.
Consider a computational
graph with $n$ outputs and $m$ inputs modeled as a function $S$ as illustrated in Figure~\ref{fig:reverse-mode-ad}, where
$S: \mathbb{R}^m \rightarrow \mathbb{R}^n$.
Let
$S({\bf x}) = (h \circ g)({\bf x})$ represent the composition of two functions
 $h$ and $g$ capturing how computational directed acyclic graphs (DAGs) are recursively processed
where
$g : \mathbb{R}^m \rightarrow \mathbb{R}^k$ maps the inputs to an intermediate space, and $h : \mathbb{R}^k \rightarrow \mathbb{R}^n$ maps the intermediate values to the outputs.

%
Then, the error in $S({\bf x})$ can be computed via reverse-mode
symbolic
differentiation by computing a Jacobian, $J$, where the
$J_{ij}$ component is given by
\begin{equation}
\label{eq:reverse-mode}
J_{ij} = \dfrac{\partial S_i}{\partial x_j}
       = \dfrac{\partial h_i}{\partial g_1} \dfrac{\partial g_1}{\partial x_j} + \dots + \dfrac{\partial h_i}{\partial g_k}\dfrac{\partial g_k}{\partial x_j} 
\end{equation}

This process efficiently computes gradients by propagating derivatives from the outputs back through the computational graph, leveraging the chain rule to accumulate contributions from intermediate variables.
Other symbolic differentiation systems such as in Machine Learning/HPC compute and accumulate concrete backward derivatives during the (so called) backward pass.
Unlike this approach,
\satire's approach is directed at
estimating the worst-case error across all input intervals, and
proceeds as follows, for all
nodes $S_j$ shown (Figure~\ref{fig:reverse-mode-ad}):
(1)~we symbolically accumulate the  derivatives from node $S_n$
walking back to node $S_j$;
(2)~we would have computed the forward symbolic expression at node $S_j$; we multiply it with $\mathbf{u}$  to get the symbolic error at $j$;
(3)~multiply the derivatives and symbolic error and accumulate the error directly at $S_n$---treating it as
a conservative estimate of
$S_j$'s contribution
to the output $S_n$ (we
sum over
all the $S_j$ nodes in the 
backward pass, as elaborated in \S\ref{sec:psr});
(4)~pass the accumulated
error expression at $S_n$
to a global optimizer that finds the maximum error at $S_j$ across all input domains $I_1$ through $I_m$.
This is how Equation~\ref{eq:total-final-form} is realized in \satire (we follow this style of derivation first introduced in \cite{darulova-phd}).

\section{Overview}
\label{sec:overview}

Figure~\ref{fig:overview} provides an overview of the various stages involved in the execution of \satire. The process begins with \satire reading a problem definition file written in a simple syntax that describes expressions and input intervals. This file may also include optional command-line arguments specifying abstraction parameters and instability control parameters. Each symbolic variable is derived from the {\em var} datatype of {\tt SymEngine}~\cite{symengine}.

Initially, \satire constructs the full abstract syntax-tree (AST) based on the input. If abstraction is disabled, \satire directly attempts to solve the complete error expression. Otherwise, it enters the abstraction loop using default abstraction parameters, which can be overridden by the user. During abstraction, all nodes at the heuristically determined abstraction height h (\S\ref{sec:abstractions}) are abstracted. For the selected candidate nodes, the error expression is computed. Once abstraction is complete, the abstracted nodes are treated as free variables with associated concrete function and error intervals. Subsequently, an AST with its height reduced by h is reconstructed. If necessary, the abstraction process continues for the remaining AST. Otherwise, the full error expression is solved for the residual AST.


Next, \satire invokes an expression builder to assign symbolic expressions to each node in the AST, performing expression canonicalization (\S\ref{sec:canonicalize}) using the {\em expand} functionality of {\tt SymEngine}---the expression simplifier that underlies \satire. The Expression Builder then performs a depth-first traversal starting from the root nodes. For multi-rooted expression DAGs, \satire supports solving them by processing all roots. This is followed by a breadth-first traversal to compute the {\em reverse-mode symbolic derivatives}, a functionality specifically built as part of \satire. Once both the symbolic expression and its derivative are computed for a node, \satire generates the corresponding symbolic error expression contributed by the node to each DAG output. These error expressions are aggregated in accumulators maintained at the output nodes. The derivative evaluation and error generation processes are synchronized to minimize multiple traversals of the AST. When the error accumulation at the output nodes is complete, the results are passed to the global optimizer (Gelpia~\cite{fptaylor,gelpia-github}) to determine concrete error bounds.

\section{Path Strength Reduction}
\label{sec:psr}

\begin{figure}[b]
\centering
    \includegraphics[width=0.5\textwidth]{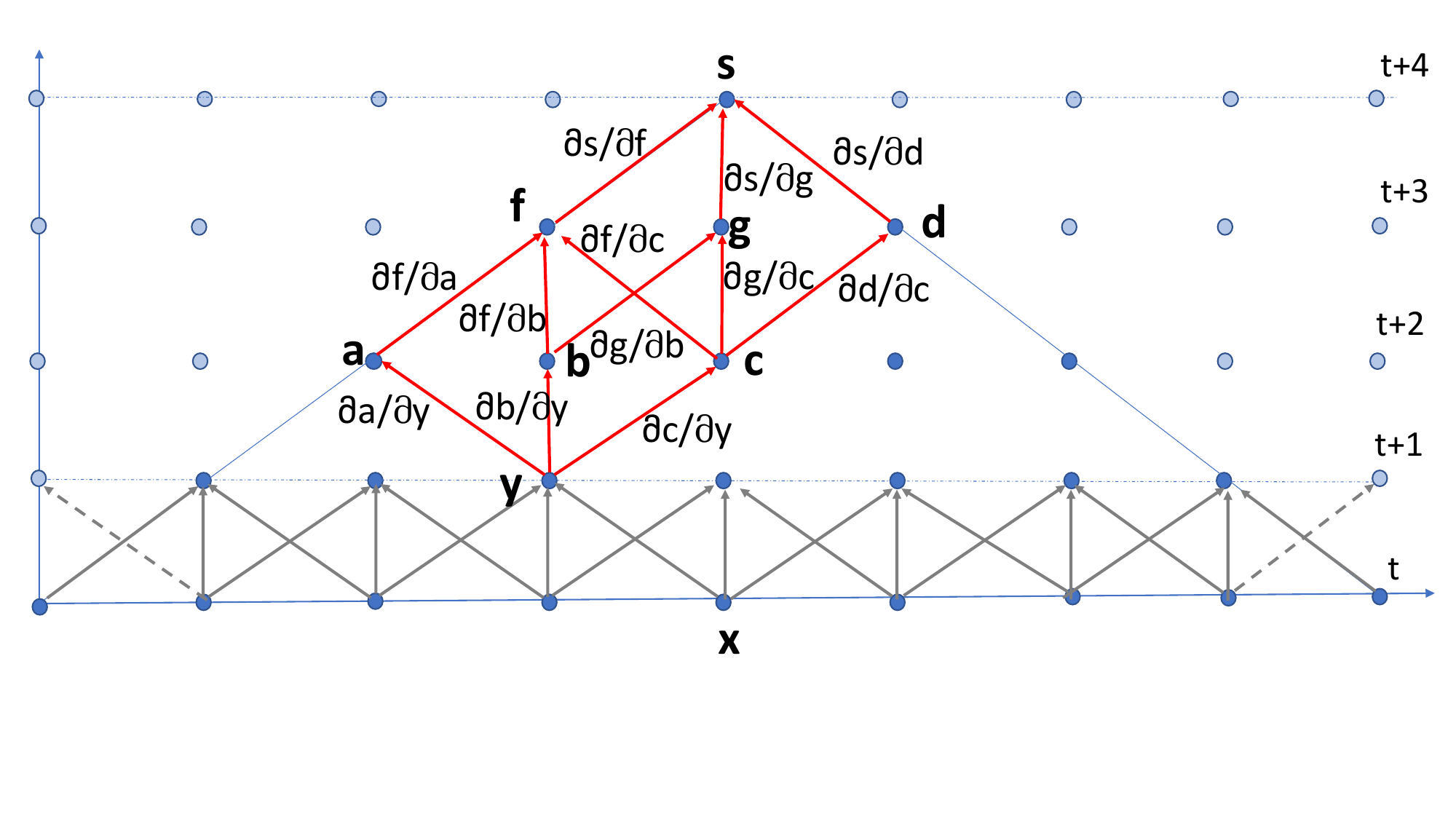}
\caption{Path strength information for  error propagating from {\bf y} to {\bf s} in an unrolled 3-point stencil evaluation across five time steps }
\label{fig:psr-example}
\end{figure}

We illustrate the concept of path strength reduction using the example of a simple unfolded 3-point stencil, as shown in Figure~\ref{fig:psr-example}. Consider the task of determining the contribution of the local error ${\mathcal E}^{lr}(y)$, introduced at node $y$, to the value at node $s$ located at $(x,t+4)$.
For the types of examples we analyze, ${\mathcal E}^{lr}(y)$ typically manifests as a very large symbolic expression. This expression is derived by performing forward symbolic execution from the primary inputs to compute the symbolic value of $y$ and then applying Equation~\eqref{eq:floating-point-operation} to scale this value by $\delta$. 

An efficient method to determine the contribution of ${\mathcal E}^{lr}(y)$ to the value at node $s$, without enumerating all paths, involves decoupling the error terms from their propagation effects. Specifically, this approach first summarizes all symbolic partial path strengths from $s$ to $y$ and then computes their product with the symbolic error term generated at $y$, namely ${\mathcal E}^{lr}(y)$. Here, path strength refers to the amplification of an input change along a given path. \satire employs a dynamic-programming-based symbolic reverse mode algorithmic differentiation to efficiently perform this calculation. This method is economical, as the derivative for each parent node is computed before reaching its child nodes, thereby sharing the derivative computation cost across all nodes. As illustrated in Figure~\ref{fig:psr-example}, the process begins by computing the partial derivatives for the immediate children of $s$. The path strength information is then propagated iteratively along their descendants in a reduction process. Additionally, symbolic expression canonicalization, as discussed in \S\ref{sec:canonicalize}, is applied throughout this process to maintain simplified and canonicalized expressions. This ensures both computational efficiency and clarity in the resulting symbolic expressions.

To compute the complete path strength from $y$ to $s$, denoted as $ps_{y\rightarrow s} = ds/dy$, we evaluate the following expression:

\begingroup
\small
\begin{equation}
    \begin{split}
        ps_{y\rightarrow s} =  ps_{y\rightarrow{a}}\cdot ps_{a\rightarrow{s}} +
                                      ps_{y\rightarrow{b}}\cdot ps_{b\rightarrow{s}} +
                                      ps_{y\rightarrow{c}}\cdot ps_{c\rightarrow{s}} 
      \implies \dfrac{ds}{dy} = \dfrac{\partial a}{\partial y}\dfrac{\partial s}{\partial a} +
                         \dfrac{\partial b}{\partial y}\dfrac{\partial s}{\partial b} +
                         \dfrac{\partial c}{\partial y}\dfrac{\partial s}{\partial c}
    \end{split}
\end{equation}
\endgroup

This computation is performed for every internal node $y_i$ (including error-laden inputs) within the dependence cone of $s$. The symbolic partial error expressions from all such nodes are then accumulated to derive the global total error expression, ${\mathcal E}^{tr}(s)$, where ${\bf y}$ represents the set of all nodes within the dependence set of $s$
: ${\mathcal E}^{tr}(s) = \sum_{y_i \in {\bf y}} {\mathcal E}^{tr}(s|y_i)$.
%
It is important to note that the path strength summaries scale linearly with the number of nodes, rather than exponentially with the number of paths. This linear scalability significantly improves performance. 


\section{Error-Bound Optimization}
\label{sec:canonicalize}

The goal of concrete error-bound calculation (Equation~\eqref{eq:total-final-form}) is to maximize the error expression over specified input intervals. Specifically, given an $n$-variable total error expression ${\mathcal E}^{tr}(s_n)$, the Interval Branch and Bound Analysis (IBBA) method~\cite{alliot2012finding} can search an $n$-dimensional box of input intervals. During each IBBA step, the initial $n$-dimensional box is recursively subdivided (within a predefined limit on the number of subdivisions) into smaller sub-boxes. Each sub-box is then queried to compute the maximum error-bound within that interval. This recursion terminates at a primitive interval library, such as Gaol~\cite{gaol}, which is part of the Gelpia optimizer~\cite{gelpia-github}. Gaol produces output bounds for these primitive intervals, and the final result is the supremum over all $n$-dimensional sub-boxes, providing the tightest error upper-bound within a specified tolerance. If ${\mathcal E}^{tr}(s_n)$ is expressed with repeated occurrences of the same variable, the computed bounds can become overly inflated. For example, consider the bound for $x\cdot x \cdot x$ with $x \in [-1,5]$. If evaluated as successive interval multiplications, Gaol produces the interval $[-25,125]$. However, by expressing this query as $pow(x,3)$, which accounts for variable sharing, the tighter bound $[-1,125]$ is obtained.

Unfortunately, Gaol cannot automatically handle all such cases using specialized functions like $pow$. To mitigate this, users must pre-process the expressions by reassociating common coefficients and grouping correlated terms before invoking Gaol. This can be achieved using tools like {\tt SymEngine}~\cite{symengine} to perform canonicalizations and simplifications.
As an example,
consider the `Direct Quadrature Moments Method’ (DQMOM) benchmark Equation$~\eqref{eq:dqmom}$,
where $I(m_i) = [-1.0,1.0]$ and parameters $w_i, a_i \in [0.00001, 1.0]$:
\begingroup
\begin{equation}
\label{eq:dqmom}
\begin{split}
	r =& (0.0 + ((((w_2 * (0.0 - m_2)) * (-3.0 * ((1.0 * (a_2/w_2)) * (a_2/w_2)))) * 1.0) + ((((w_1 * (0.0 - m_1)) * (-3.0 * ((1.0 \\
	   & * (a_1/w_1)) * (a_1/w_1)))) * 1.0) + ((((w_0 * (0.0 - m_0)) * (-3.0 * ((1.0 * (a_0/w_0)) * (a_0/w_0)))) * 1.0) + 0.0))))
\end{split}
\end{equation}
\endgroup

The non-canonicalized form of $r$ produces an interval bound of \tb{\bf [-4.5e+10, 4.5e+10]}.
After canonicalization, the simplified equivalent form, $r = 3.0*(a_0^2)*m_0/w_0 + 3.0*(a_1^2)*m_1/w_1 + 3.0*(a_2^2)*m_2/w_2$,
%
%
which noticeably reduces the
distinct occurrences of $w_i$ and $a_i$, reduces the interval bound to {\bf [-9.0e+05, 9.0e+05]} - 
a five orders of magnitude improvement.
Using this improved interval, \satire reports an error-bound of \tb{\bf 5.0e-10}
for DQMOM, compared to \tb{\bf  3.45e-05} reported by FPTaylor, which does not perform canonicalization. A shadow-value computation confirms \satire's result with a bound of {\bf  3.27e-13}, further demonstrating its bound tightness. It is important to note that canonicalization simplifies only the query expression submitted to Gaol; it does not alter the abstract syntax-tree (AST) being analyzed, ensuring that floating-point error analysis remains unaffected.

%

The combination of path strength reduction and canonicalization allows \satire to outperform FPTaylor significantly (refer \cite{2020_Das}, often handling expressions with up to 10K operators without requiring abstraction. Furthermore, \satire achieves tight bounds on smaller benchmarks while being, on average, 4.5x faster. These advantages are further amplified when applying abstractions (\S\ref{sec:abstractions}) to handle larger examples (\S\ref{sec:eval}) with similar benefits. However, type-based analysis systems like $\Lambda_{num}$ do seem to perform faster and provide tighter bounds in cases presented in \cite{2025_ariel}.

%


\section{Computation Graph Abstractions}
\label{sec:abstractions}

\begin{figure*}[htbp]
	\centering
    \begin{minipage}{0.43\textwidth}
    \includegraphics[width=0.85\columnwidth]
    {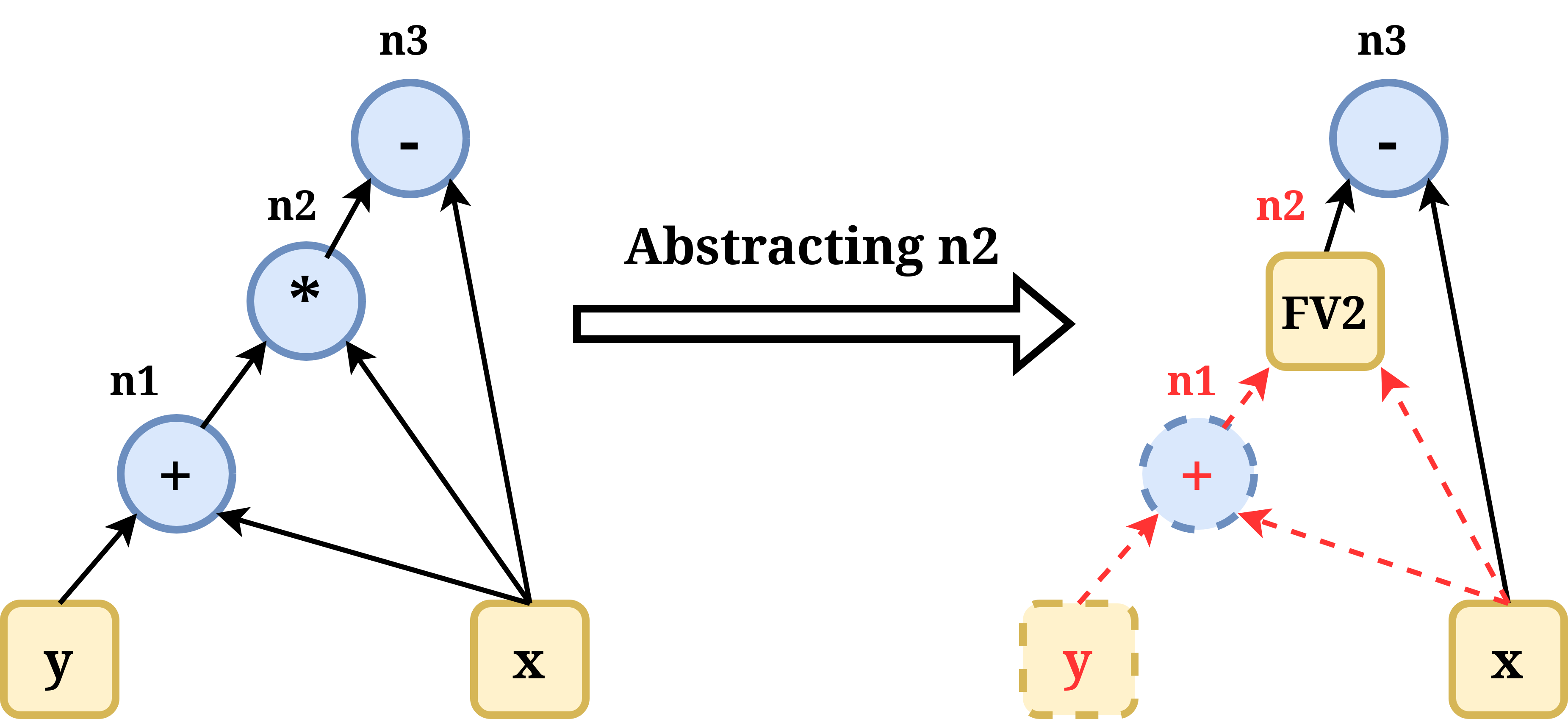}
  \caption{\label{fig:abt-abstraction}Abstractions introduced in a simple expression AST}
  \end{minipage}
  \hfill
  \begin{minipage}{0.43\textwidth}
   \includegraphics[width=1.05\columnwidth]{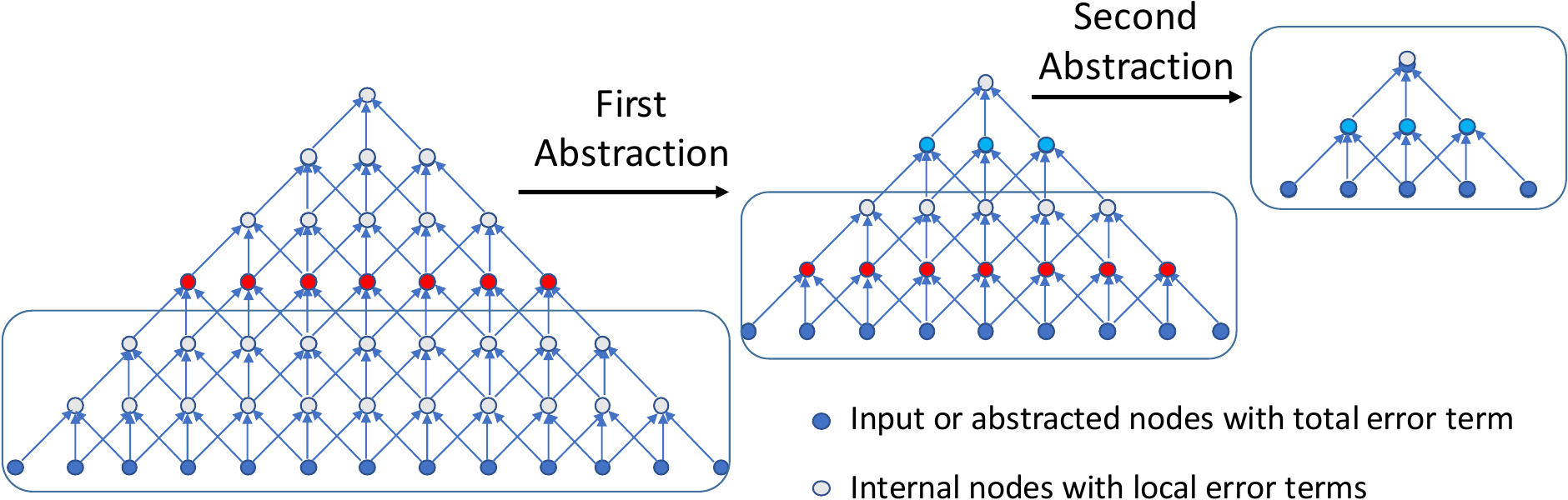}
  \caption{\label{fig:gradual-abstraction}Incremental error analysis using gradual abstraction}
  \end{minipage}
\end{figure*}

A central strategy for scalability in \satire is replacing entire sub-expression directed acyclic graphs (DAGs) with their root node, which summarizes the error up to that point. In Figure~\ref{fig:abt-abstraction}, a newly introduced input node (a ``free variable'' $FV2$) replaces the DAG under the `*' operator. For the remainder of the analysis, $FV2$ carries the concrete error ${\mathcal E}^{lr}(n2)$, instead of the symbolic error associated with $n2$.\footnote{The concrete error $E^{tr}(FV2)$ is computed using the global optimizer, as described in Equation~\eqref{eq:total-final-form}.}

This approach offers two key advantages:(1) Large portions of the expression graph, along with their symbolic errors, are excluded from further analysis, reducing computational burden. (2) Concrete errors, treated as constants, can be seamlessly incorporated into subsequent symbolic analysis, further improving efficiency.\footnote{\satire combines interval and symbolic-affine analysis methods, rather than relying solely on interval-based or affine-based approaches.} However, excessive use of abstractions can introduce challenges. Figure~\ref{fig:abt-abstraction} demonstrates how inputs such as $x$ can propagate to both the abstracted node and ``higher up'' in the graph. This creates two primary disadvantages: (1) Loss of variable correlation: When variables become uncorrelated, the global optimizer may overestimate errors, as discussed in \S\ref{sec:canonicalize}. For instance, with $I(x)=I(y)=[-1,1]$, abstraction results in $FV2 = [-2.0, 2.0]$, and the output of the `$-$' node $(FV2 - x)$ evaluates to $[-3,3]$. Without abstraction, the output of the `$-$' node in $(x+y)*x - x$ evaluates to the tighter interval $[-2,3]$. (2) Missed opportunities for error cancellation: Error cancellation is crucial in floating-point error analysis, and abstraction can eliminate these opportunities, leading to less precise error bounds.
    
The abstraction heuristic used in \satire based on the notion of entropy is detailed in \cite{2020_Das}. This heuristic    aims to strike a practical balance between several competing factors that influence its effectiveness, while remaining fast and straightforward to implement. The efficacy of abstraction depends on various factors, including expression sizes, the degree of non-linearity, the extent of variable correlation loss, and the execution time for each optimizer call.

\section{Handling Conditionals}
\label{sec:error-accumulation}

\begin{figure*}[htbp]
	\centering
    \begin{minipage}{0.48\textwidth}
\begin{lstlisting}[
    caption=\satire Program, 
    captionpos=b,
    numbers=left,
    breaklines=true,
    basicstyle=\small,
    numberstyle=\small\color{gray},
    xleftmargin=0.75cm,
    numbersep=5pt,
    label={satire_input}]
 INPUTS {x fl64 : (-1.0, 1.0);
         y fl64 : (-1.0, 1.0);}

 OUTPUTS {res;}

 EXPRS { h = (y+x)*x;
         g = x+x*y
         if (x-y < 0.4) then
             g = 1+1/g;
         else
             if (x*x>0.25) && 
             (y*h <= x*x) then
                 g = h-x;
             endif
         endif 
         res = g+y;}
\end{lstlisting}
  \end{minipage}
  \hfill
  \begin{minipage}{0.48\textwidth}
   \includegraphics[width=0.75\linewidth]{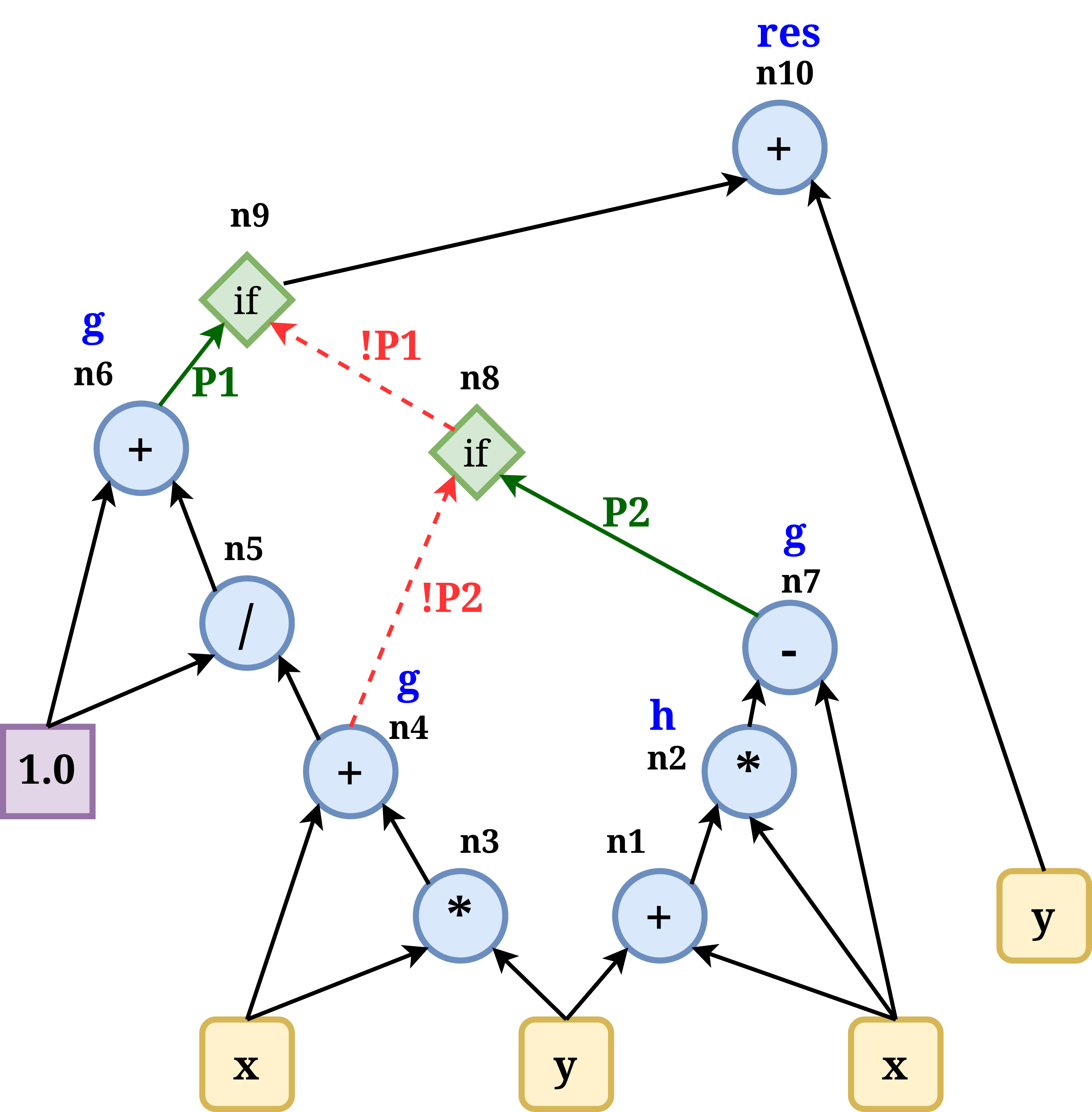}
    \caption{Conditional Computation Graph}
    \label{fig:CCG}
  \end{minipage}
\end{figure*}



In \satire, we extend the efficient dynamic-programming implementation of this approach to handle conditional paths, as detailed in~\cite{2021_Das}. \satire introduces key methodological advancements, including the evaluation of partial derivatives over conditional expressions, the separation of conditional predicates for instability analysis, and the maximization of error expressions over domains constrained by predicate expressions, as now detailed.


\subsection{\em Local Error Analysis} 

Consider the \satire format input program in Listing~\ref{satire_input} and its corresponding Computation graph displayed in Figure~\ref{fig:CCG}. Look at how the local error generated at node $\mathbf{n3}$, represented as \( \mathcal{E}^{lr}(\mathbf{n3}) = (x\cdot y)\cdot \mathbf{u} \), propagates to the output node $res$. This local error term propagates to the output along two paths: $\sigma_1 : (\mathbf{n3},\mathbf{n4},\mathbf{n5},\mathbf{n6},\mathbf{n9},\mathbf{res})$, and $\sigma_2 : (\mathbf{n3},\mathbf{n4},\mathbf{n8},\mathbf{n9},\mathbf{res})$.
    
Let the worst-case impact of $\mathcal{E}^{lr}(\mathbf{n3})$ on $\mathbf{res}$ be denoted as $\mathcal{E}^{tr}(\mathbf{res} | \mathbf{n3})$. To compute this, we evaluate the partial derivative of $\mathbf{res}$ with respect to $\mathbf{n3}$ along each of these paths and add the results. This approach provides a conservative over-approximation, ignoring error cancellation effects that cannot be modeled in symbolic analysis. These derivatives describe the path strength for error propagation. However, to ensure precision, we must derive path-specific derivatives modeled as predicated expressions that determine whether the conditional paths are "active." Using the notation $[\![P\rightarrow E]\!]$, we represent the expression ``if $P$ then $E$.'' In more detail, for path $\sigma_1$, we apply reverse-mode differentiation from the output $\mathbf{res}$ back to $\mathbf{n3}$, as shown in Equation~\eqref{eq:deriv-path1}:

\begingroup
\begin{equation}
\label{eq:deriv-path1}
    \begin{split}
        \frac{\partial res}{\partial n3} \bigg |_{\sigma_1}
        &= \frac{\partial res}{\partial n9}\cdot \frac{\partial n9}{\partial n6}\cdot
                   \frac{\partial n6}{\partial n5}\cdot\frac{\partial n5}{\partial n4}\cdot\frac{\partial n4}{\partial n3} \\
        &= [\![T \rightarrow 1]\!] \cdot [\![P1 \rightarrow 1]\!]\cdot [\![T \rightarrow 1]\!]\cdot [\![T \rightarrow \frac{-1}{(x+xy)^2}]\!]\cdot[\![T \rightarrow 1]\!] \\
        &= [\![P1 \rightarrow \frac{-1}{(x+xy)^2}]\!]
    \end{split}
\end{equation}
\endgroup

Here, $T$ denotes  ``True''. The derivative along path $\sigma_2$ is similarly computed as $\frac{\partial res}{\partial n3} \bigg |_{\sigma_2} = [\![(\lnot P1 \wedge \lnot P2) \rightarrow 1]\!]$. It is important to note that these two paths are mutually exclusive only assuming that the predicates are rounding-error free. The next section addresses
how the predicate expression errors create the instability windows.
%

\subsection{\em Instability Analysis and Predicate Weakening} 

  When rounding errors propagate into two otherwise mutually exclusive predicates, $P$ and $\neg P$, the predicates develop an ``overlap region.'' This overlap is modeled by {\em weakening} (rewriting, denoted by superscript ``w'') the predicates to $P^w$ and $(\neg P)^w$. Intuitively, weakening transforms predicate expressions, such as $E_1 < E_2$, into $(E_1 - \mathcal{E}^{tr}_{E_1}) < (E_2 + \mathcal{E}^{tr}_{E_2})$, where $\mathcal{E}^{tr}_{E_1}$ and $\mathcal{E}^{tr}_{E_2}$ represent the rounding errors associated with $E_1$ and $E_2$, respectively. This models the effect of rounding errors on the conditional expressions. The precise definition of weakening during error analysis is involved,
  and is formally detailed in~\cite{2021_Das} as a case analysis over the syntax of predicate expressions and also previously presented in Rosa~\cite{rosa} and \precisa~\cite{precisa}. After weakening, the conjunction $P^w \wedge (\neg P)^w$ is no longer false but becomes satisfiable within the overlap region of the ``error-infused'' predicate expressions.

\subsection{\em Conditional Accumulation}

To model accumulation over predicated paths, we introduce a new conditional accumulation operator, denoted as $\bowtie$, formally defined as follows (it helps recursively fuse the overlap regions):
    \vspace{-.9ex}
    \begin{equation}
    \label{eq:basic-bowtie}
        \begin{split}
            [\![Pred_i \rightarrow PExpr_i]\!] \bowtie [\![Pred_j \rightarrow PExpr_j]\!] = & [\![(Pred_i \wedge Pred_j) \rightarrow PExpr_i \bowtie PExpr_j \\
            &\ |\ \  (Pred_i \wedge \lnot Pred_j) \rightarrow PExpr_i \\
            &\ |\ \  (\lnot Pred_i \wedge Pred_j) \rightarrow PExpr_j \\
            &\ |\ \  (\lnot Pred_i \wedge \lnot Pred_j) \rightarrow 0]\!]
        \end{split}
    \end{equation}

The $\bowtie$ operator partitions the solution domain based on the constraints defined by the predicates. It takes the Boolean combinations of the predicates. It evaluates $PExpr_i \bowtie PExpr_j$ where $Pred_i$ and $Pred_j$ overlap. For points where $Pred_i$ holds but $Pred_j$ does not, it selects $PExpr_i$. Conversely, for points where $Pred_j$ holds but $Pred_i$ does not, it selects $PExpr_j$. For points outside both predicates, the operator assigns $0$, modeling the absence of a path contribution.

In practical scenarios, $Pred_i$ and $Pred_j$ often originate as mutually exclusive branches of an "if-then-else" construct, modeling real-valued semantics. However, under floating-point semantics, predicate weakening introduces overlap between $Pred_i$ and $Pred_j$, making all cases of $\bowtie$ relevant. This overlap creates a ``sneak path'' in which both the ``then'' and ``else'' branches can contribute to the output within the overlap region but we do not know which. Consequently, the errors accumulated in the then-branch and else-branch compared and the greater one is considered in this region. This behavior provides a conservative over-approximation of how ``if-then-else'' operates under floating-point semantics.

With these modifications, we can express $\mathcal{E}^{tr}(\mathbf{res}|\mathbf{n3})$ as follows, capturing the cumulative effects of conditional path overlaps and the corresponding contributions to error accumulation (experimental results are in 
Table~\ref{tab:seesaw-solver-compare}):
%
\begin{equation}
    \label{eq:example-bowtie}
    \begin{split}
       \mathcal{E}^{tr}(\mathbf{res}|\mathbf{n3}) = \mathcal{E}^{lr}(\mathbf{n3}) \bigg (\frac{\partial res}{\partial n3}\bigg |_{\sigma_1} \bowtie \frac{\partial res}{\partial n3}\bigg |_{\sigma_2} \bigg ) &= (xy)\cdot \boldsymbol{u} \cdot \bigg ( [\![P1^w \rightarrow \frac{-1}{(x+xy)^2}]\!] \bowtie [\![(\lnot P1)^w \wedge (\lnot P2)^w \rightarrow 1]\!] \bigg) \\
        &= [\![(P1^w \wedge (\lnot P1)^w \wedge (\lnot P2)^w) \rightarrow xy\left(1 - \frac{1}{(x+xy)^2}\right)\cdot \boldsymbol{u}]\!] \\
        &\quad\ |\ \ [\![(\lnot P1)^w \wedge (\lnot P2)^w \rightarrow - \frac{xy}{(x+xy)^2}\cdot \boldsymbol{u}]\!] \\
        &\quad\ |\ \ [\![P1^w \wedge (P1^w \vee P2^w) \rightarrow (xy)\cdot \boldsymbol{u}]\!]
    \end{split}
\end{equation}

\section{Mixed-Precision Extension}
\label{sec:mixed-precision}

In this section we will elaborate on extending the symbolic Taylor forms for mixed-precision\footnotemark programs by incorporating the rounding error incurred due to type-cast operations; here type-cast refers to the act of interfacing two precision regimes.
We apply this new model to bound the absolute error at the output.
\footnotetext{We use notations from \fptuner~\cite{fptuner} in sections related to mixed-precision.}

\subsection{Extending Formalism}

The basic idea of handling mixed-precision is to recognize that
in addition to the normal rounding
 rules, one must also ``round'' when a higher precision expression is used in a lower precision context (but not vice-versa)---called {\em type-cast} because that is akin to assigning an FP64 variable (for instance) to an FP32 variable after an explicit type-cast.
 We now describe how this strategy
 is incorporated into \satire. 
 


\begin{wrapfigure}{r}{0.3\textwidth}
    \centering
    \includegraphics[width=0.75\linewidth]{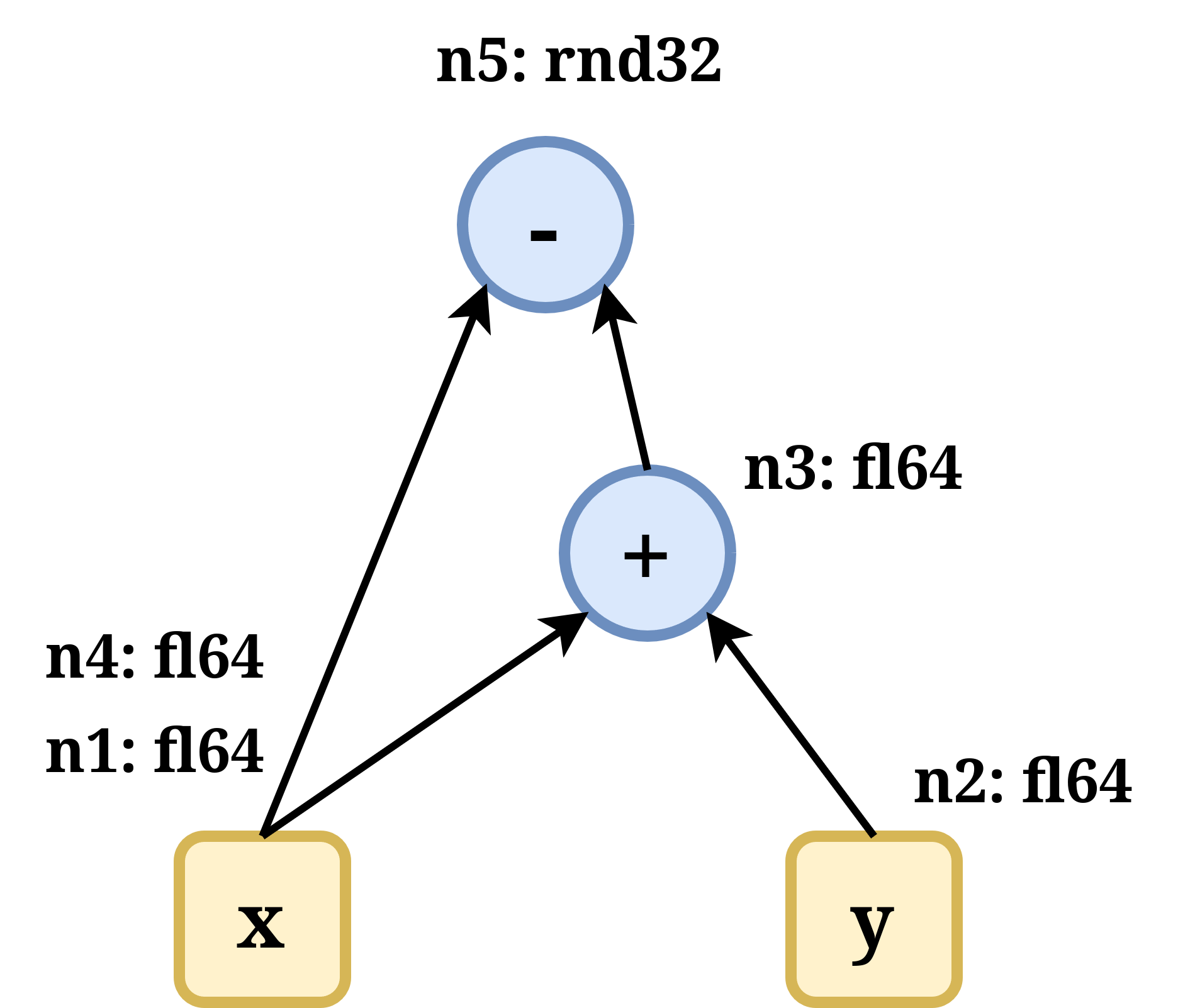}
    \caption{Mixed Precision Example}
    \label{fig:mixed_precision_example}
    
\end{wrapfigure}

We assume that when an operator yields an output of a certain precision, its inputs are set to the same precision before the operation is performed. Consider the \textbf{n5} node (Figure~\ref{fig:mixed_precision_example}) that is
set to single-precision where \textbf{n3} is set to double-precision. In this example, 
a type-cast operation is needed to
down-cast the input \textbf{n3} from double-precision to single-precision.

The `-` operator thus forms the context for its inputs \textbf{x} and \textbf{n3}.
This ``context'' determines the kind of type-casting introduced. In general, a rounding-term is added when down-casting and it is 0 when up-casting.

Each operator with post-order number \textbf{\textit{i}} is associated with two noise variables, $\delta_i$ and $\Delta_i$. We associate the two noise variables with each operator through two maps, namely $A_{\delta_i}$ and $A_{\Delta_i}$, that specify the bounds of these noise variables. More specifically, 
$A_{\delta_i}$ (resp., $A_{\Delta_i}$) yields the bound of $\delta_i$ (resp., $\Delta_i$). %
%

Consider the example expression in Figure~\ref{fig:mixed_precision_example} where $\mathscr{E} = x-(x+y)$. We have 
$A_\delta=\{(\delta_1, \epsilon_{64}), (\delta_2, \epsilon_{64}), (\delta_3, \epsilon_{64}), (\delta_4, \epsilon_{64}), (\delta_5, \epsilon_{32})\}$ and the inferred $A_\Delta=\{(\Delta_1, 0), (\Delta_2, 0), (\Delta_3, \epsilon_{32}), (\Delta_4, \epsilon_{32}), (\Delta_5, 0)\}$. The modeling expression then will be:
\begingroup
\small
\begin{equation}
\Tilde{\mathscr{E}}=(x.(1+\delta_1).(1+\Delta_4)-(x.(1+\delta_1).(1+\Delta_1)+y.(1+\delta_2).(1+\Delta_2))(1+\delta_3).(1+\Delta_3))(1+\delta_5).(1+\Delta_5)
\label{eq:modelling_expression_example}
\end{equation}
\endgroup


\subsection{Bound Determination}

We introduce the additional error term $\mathcal{T}(\mathscr{S}) = (\Tilde{x} \diamond \Tilde{y})\Delta$ representing the amount of round-off error due to type-cast operations. This additional error term is introduced in Equation$~\eqref{eq:taylor-expansion}$ 
to get a new error model for the program output in \S\ref{subsec:taylor_expansion} as follows:
\begingroup
\small
\begin{equation}
\label{eq:mixed-precision-taylor-expansion}
   \mathcal{E}^{tr}_{\mathscr{E}_{n}} = \mathscr{E}_n\delta_n
            + \sum_{j=1}^{n-1} \dfrac{\partial \mathscr{E}_n}{\partial \mathscr{S}_j}(\mathcal{E}^{lr}_{\mathscr{S}_j} + \mathcal{T}_{\mathscr{S}_j})
            + \sum_{j=1}^{m} \dfrac{\partial \mathscr{E}_n}{\partial x_j}(\mathcal{E}^{lr}_{x_j} + \mathcal{T}_{x_j})
            + O({\bf u}^2)
\end{equation}
\endgroup

From Equation$~\eqref{eq:mixed-precision-taylor-expansion}$, we can observe that $\mathcal{E}^{tr}_{\mathscr{E}_n}$ is composed of the {\em local error} generated by the application of $\dot{\mathscr{E}_n}$, which we write ${\mathcal E}^{lr}_{\mathscr{E}_n}$, and the propagation of the incoming errors $\mathcal{E}^{lr}_{x_j}$, $\mathcal{T}_{x_j}$, $\mathcal{E}^{lr}_{\mathscr{S}_j}$ and $\mathcal{T}_{\mathscr{S}_j}$~\cite{rosa,higham}.

Similar to the process in \S\ref{subsec:taylor_expansion} we can bound the error on $\mathcal{E}^{tr}_{\mathscr{E}_n}$ in Equation$~\eqref{eq:mixed-precision-taylor-expansion}$
as follows:

\begingroup
\small
\begin{equation}
\label{eq:mixed-precision-total-final-form}
\begin{split}
E^{tr}(\mathscr{E}_n) &\leq \max_{{\bf x}\in I({\bf x})}(|\mathcal{E}^{tr}_{\mathscr{E}_n}|)  \\
	 &\leq \max_{{\bf x}\in I({\bf x})}
         \bigg (|\mathcal{E}^{lr}_{\mathscr{E}_n}| + |\sum_{j=1}^{n-1} \dfrac{\partial \mathscr{E}_n}{\partial \mathscr{S}_j}
         (\mathcal{E}^{lr}_{\mathscr{S}_j} + \mathcal{T}_{\mathscr{S}_j})| + |\sum_{j=1}^{m} \dfrac{\partial \mathscr{E}_n}{\partial x_j}(\mathcal{E}^{lr}_{x_j} + \mathcal{T}_{x_j}) | 
					\bigg ) + O({\bf u}^2)
\end{split}
\end{equation}
\endgroup

Note that $\mathscr{E}_n$ does not have a type-cast rounding term since it depends on the context and $\mathscr{E}_n$ has none so it is assumed to be 0.

\section{Evaluation}
\label{sec:eval}



\paragraph{{\bf Experimental Setup: }}
All benchmarks were executed with Python3.8.0 version on a
dual 14-core Intel Xeon CPU E5-2680v4 2.60GHz CPUs system (total
28 processor cores) with 128GB of RAM running Linux OS.
\satire is compatible with Python3. The symbolic engine is based on {\tt SymEngine}~\cite{symengine}. It also depends on Gelpia~\cite{gelpia-github} as its global optimizer.
The core analysis algorithms were measured without any multicore parallelism although \satire supports it.

During this evaluation we will be re-running larger benchmarks to test scalability and conditional benchmarks to test for any changes due to our mixed-precision extensions. Note that these are instantiated using the double-precision floating-point type. However, our main focus will be generating uniform single-precision and mixed single,double-precision benchmarks to test the mixed-precision extension for performance and accuracy in \S\ref{sec:mix-prec-eval}. We also run a case study on Conjugate Gradient by increasing precision of key computation steps and compare this accuracy with a uniform precision version.

Our larger benchmarks encompass a diverse range of computational problems, including 
stencils (e.g., Finite-Difference Time-Domain (FDTD)), 
iterative solvers, 
Fast Fourier Transforms (FFT), 
matrix multiplication
.
We adapted stencil kernels for 
heat (H*), 
Poisson (P*), 
and convection-diffusion (C) types from~\cite{fpd}, 
unrolling them over 32 steps 
to conform to \satire's input format. 
%
%
For the Conjugate Gradient (CG) solvers, 
input matrices were sourced from~\cite{ufl} 
and analyzed over 20 iterations of the solver.

\begingroup
\small
 \begin{table*}[ht]
 \centering
 \begin{threeparttable}[t]
\setlength\tabcolsep{3.0pt}
  \begin{tabular}{crrrrrrrrr}
  \toprule
  \multirow{2}{*}{Benchmarks} & \multirow{2}{*}{\shortstack[1]{Num \\OPs \\ (in \\1000's)}} & \multicolumn{2}{c}{Direct Solve} & \multicolumn{2}{c}{Window-(10,20)} & \multicolumn{2}{c}{Window-(15,25)} & \multicolumn{2}{c}{Window-(20,40)} \\

  \cmidrule(lr){3-4}  \cmidrule(lr){5-6}  \cmidrule(lr){7-8}  \cmidrule(lr){9-10}

  & & \shortstack[1]{Absolute\\Error\\Bound} & \shortstack[1]
  {Execution \\Time\\ (secs)} & \shortstack[1]{Absolute\\Error\\Bound} & \shortstack[1]{Execution \\Time\\(secs)} & \shortstack[1]{Absolute\\Error\\Bound} & \shortstack[1]{Execution \\Time\\(secs)} & \shortstack[1]{Absolute\\Error\\Bound} & \shortstack[1]{Execution \\Time \\ (secs)} \\
  \midrule\\
  {Scan(1024pt)} & 2 & \textbf{9.38e-13} & \textbf{169}  & \textbf{9.38e-13} & \textbf{173}  & \textbf{9.38e-13} & \textbf{175}  & \textbf{9.38e-13} & \textbf{175}  \\
  {Scan(4096pt)} & 8 & \textbf{4.66e-11} & 3387  & \textbf{4.66e-11} & \textbf{13}  & \textbf{4.66e-11} & 3401  & \textbf{4.66e-11} & 3427  \\
  {FFT-1024pt} & 81 & \textbf{3.98e-13} & 205 & 4.52e-13 & \textbf{19} & 4.36e-13 & 30 & \textbf{3.98e-13} & 188 \\
  {FFT-4096pt} & 393 & \textbf{1.82e-12} & 2839 & 2.02e-12 & \textbf{74} & 1.935e-12 & 126 & \textbf{1.82e-12} & 2833 \\
   H1\tnote{a} & 393 & NA & NA & \textbf{1.11e-14} & \textbf{1454} & \textbf{1.11e-14} & 3805& \textbf{1.11e-14} & 11376 \\
   H2\tnote{a} & 393 & NA & NA & \textbf{1.94e-14} & \textbf{1474} & \textbf{1.94e-14} & 3885& \textbf{1.94e-14} & 11598\\
   H0\tnote{a} & 393 & NA & NA & \textbf{5.55e-14} & \textbf{1467} & \textbf{5.55e-14} & 4162& \textbf{5.55e-14} & 11542\\
   P1\tnote{a} & 436 & NA & NA & \textbf{2.26e-14} & \textbf{757} & \textbf{2.26e-14} & 1339 & \textbf{2.26e-14} &  2780 \\
   P2\tnote{a} & 436 & NA & NA & \textbf{2.26e-14} & \textbf{674} & \textbf{2.26e-14} & 1335 & \textbf{2.26e-14} &  2670 \\
   P0\tnote{a} & 436 & NA & NA & \textbf{7.55e-14} & \textbf{667} & \textbf{7.55e-14} & 1478 & \textbf{7.55e-14} & 2639 \\
   C1\tnote{a} & 436 & NA & NA & \textbf{6.25e-15} & \textbf{1950} & \textbf{6.25e-15} & 4315 & \textbf{6.25e-15} & 11502 \\
   C2\tnote{a} & 436 & NA & NA & \textbf{6.25e-15} & \textbf{1287} & \textbf{6.25e-15} & 4358 & \textbf{6.25e-15} & 11840 \\
   C0\tnote{a} & 436 & NA & NA & \textbf{5.56e-14} & \textbf{1752} & \textbf{5.56e-14} & 4580 & \textbf{5.56e-14} & 11865 \\
   Advection & 453 & \textbf{1.01e-13} & \textbf{3533} & \textbf{1.01e-13} & 3630 & \textbf{1.01e-13} & 3724 & \textbf{1.01e-13} & 3547 \\
   FDTD & 192 & \textbf{3.71e-13} & 
                             \textbf{9792} & -- & -- & -- & -- & -- & --  \\
  matmul-64 & 520 & \textbf{4.76e-13} & 78 & \textbf{4.76e-13} & 65 & \textbf{4.76e-13} & 61 & \textbf{4.76e-13} & \textbf{63} \\
  matmul-128 & 4177 & \textbf{1.86e-12} & 840 & \textbf{1.86e-12} & 682 & \textbf{1.86e-12} & 580 & \textbf{1.86e-12} & \textbf{538} \\
  {Tens contrac} & 1683 & NA & NA & \textbf{2.28e-13} & 4043 & \textbf{2.28e-13} & \textbf{3426} & \textbf{2.28e-13} & 3456 \\
  {Tens contrac}\tnote{b} & 1683 & NA & NA & 1.57e-19 & 1620 & 1.57e-19 & 1049 & 1.57e-19 & 827 \\
  {CG-Arc}\tnote{b} & 211 & \textbf{3.15e-19} & 1448 & 1.23e-17 & 2520 & 1.68e-17 & \textbf{1136} & 1.54e-17 & 4252  \\
  {CG-Pores}\tnote{b} & 45 & \textbf{2.67e-05} & \textbf{12} & -- & -- & -- & -- & -- & -- \\
  {Mol-Dyn}\tnote{c} & 7 & NA & NA & \textbf{3.96e-14} & \textbf{54} & 4.46e-14 & 135 & 4.46e-14 & 135 \\
  Serial Sum & 1 & \textbf{2.91e-11} & 5524 & \textbf{2.91e-11} & 16 & \textbf{2.91e-11} & \textbf{12} & \textbf{2.91e-11} & \textbf{12} \\
  Reduction & 1 & \textbf{5.68e-13} & \textbf{39} & \textbf{5.68e-13} & 36 & \textbf{5.68e-13} & 35 & \textbf{5.68e-13} & 36 \\
  Poly-50   & 1.3 & \textbf{3.26e-13} & 4 & 6.48e-13 & 3 & 6.25e-13 & 3 & 6.22e-13 & \textbf{2} \\
  Horner-50 & 0.1 & \textbf{1.03e-13} & 6 & 4.43e-13 & 3 & 2.14e-13 & \textbf{2} & 2.58e-13 & 2 \\
	\end{tabular}
	\begin{tablenotes}
	\item[a] Benchmarks ported from~\cite{fpd} and ~\cite{fpdetect}
	\item[b] Benchmarks analyzed over degenerate intervals
	\item[c] Benchmark realized from ~\cite{MD}
	\end{tablenotes}
	\caption{\label{tab:satire-results} Satire evaluated worst case absolute error bounds and execution time on larger benchmarks. }
 \end{threeparttable}
 \end{table*}
\endgroup


Table~\ref{tab:satire-results}\footnotemark summarizes these large benchmarks, 
where the "Direct Solve" column indicates error bounds computed without applying abstractions.
The "Num OPs" column shows the number of operators in a single expression tree, resulting from unrolling the loops in the benchmarks to various sizes. 
The unroll factors were selected with two main objectives: (1) To stress-test \satire's capabilities and scalability and, (2) To simulate practical use cases, where a user might examine error propagation across iterations to understand error buildup.

\footnotetext{We refer \cite{2025_ariel} to compare error bounds and performance numbers with $\Lambda_{num}$ and Gappa due to lack of space}


Our benchmark suite also covers examples involving floating-point expressions
impacting the control flow. Such algorithms
(common in computational geometry~\cite{compgeo}
and collision detection problems~\cite{realtime})
are typically designed under an assumption of computations being
performed in exact real arithmetic.
Given the
numerical and geometrical nature of these 
algorithms, they are particularly sensitive to 
numerical robustness problems. 
Additionally, we include benchmarks involving
convergence testing such as in partial differential equation solving,
Conjugate Gradient methods and Gram-Schmidt orthogonalization.
We have also ported additional existing benchmarks for straight-line
codes with and without branches from other tools suites such as
\precisa and \fptaylor.

\subsection{Tool scalability without conditionals and efficacy of Abstractions}

In our experiments pertaining
to abstraction, we observed the competing forces of frequent and infrequent abstractions
mentioned in \S\ref{sec:abstractions} at play.
A notable finding was that while frequent abstractions weakened the error bounds, the impact was minimal—error bounds remained within the same order of magnitude.
However, certain cases were more affected. For example, in FDTD computations, the concretization of internal nodes during abstraction introduced larger correlation losses, leading to increased error bounds. Fortunately, FDTD does not critically rely on abstractions, and in such cases, the preservation of correlations outweighed the need for abstraction, resulting in favorable error cancellations.

We experimented with three candidate abstraction windows: (10,20), (15,25), and (20,40). In most cases, smaller abstraction windows produced reasonably tight error bounds while significantly reducing execution time. This reduction in computation time was particularly beneficial when numerous optimizer queries were involved.

We now discuss the impact of semantics-preserving code transformations on abstraction. 
Since such transformations alter the expression DAG, they naturally affect the resulting round-off error bounds.
In one study, a serial summation of 1,024 points produced an error bound of 2.91e-11, 
while a reduction-tree-based summation of the same elements yielded a much tighter error bound of 5.68e-13 — 
a two order of magnitude improvement. 
Similarly, we compared standard polynomial evaluation to Horner's method, which is designed to minimize numerical error. Both Horner's method and tree reduction, as confirmed by \satire, are inherently more precision-preserving than their counterparts (standard polynomial evaluations and serial summation, respectively).
We also observed that abstraction had a negligible impact on error bounds for reduction-based schemes. For Horner's method, there was a slight loss in the tightness of error bounds: the original bound of 1.0e-13 (without abstraction) increased to 4.43e-13 with an abstraction window of (10,20).

In summary, these equivalence-preserving transformations generally maintain the same order of magnitude for error bounds, whether or not abstraction is applied. However, as expected, larger abstraction depths tended to produce tighter bounds in such scenarios.

\subsection{Evaluation for tests with conditionals}

Our evaluation of \satire
emphasizes the following aspects when analyzing floating-point programs with conditionals:

\subsubsection{\bf Worst case error bounds}

%

We report the bounds on the maximum absolute
  {\em roundoff error} that can be observed
  at the program output(s).
  This analysis does not suppress inputs that fall into
  instability regions
  (constraints are comprised of weakened predicates).\footnote{\satire also
  supports the option to suppress such inputs; keeping these inputs gives nearly
  the same---and more conservative---results.}

Table~\ref{tab:seesaw-solver-compare} summarizes the evaluation
of benchmarks involving straight-line codes with conditionals.
Across all benchmarks, the order of absolute error bounds reported are within
one order of magnitude.
\satire obtains near optimal answers with a
short execution time in most cases.

\begingroup
	\begin{table*}[ht]
	\centering
	\begin{threeparttable}[t]
	\begin{tabular}{crrrrr}
	  \toprule
	  \multirow{3}{*}{Benchmarks} & \multicolumn{3}{c}{Worst case absolute error estimation} &  \multicolumn{2}{c}{Empirical Profiling}\\
	  \cmidrule(lr){2-4} \cmidrule(lr){5-6}
	  &      {Error} & {Exec Time} & {Instability Jump} & {\textbf{Prof Error Expr}} & {\textbf{Instability jump}}\\
	\midrule 
	Barycentric\tnote{a}  	&  1.30e-15  & 136.00s   &  7.81   	& 9.34e-16   & 0.93 \\
	SymSchur\tnote{a}	&  4.13e-16  & 5.50s    &  0.42   	& 7.42e-17   & 1.48e-08 \\
	DistSinusoid\tnote{a}   &  {\bf inf} & --       &  {\bf inf} 	& 3.78e-16   & 0.05 \\
	Interpolator\tnote{c}	&  7.49e-14  & 2.68s    &  225.00    	& 2.91e-14   & 11.25 \\
	Nonlin-Interpol		&  2.24e-14  & 0.40s    &  101.00    	& 3.22e-16   & 2.9 \\
	EnclosingSp\tnote{a}  	&  4.41e-15  & 8.01s    &  12.63  	& 1.41e-15   & 4.74e-07 \\
	cubicspline\tnote{c}	&  8.88e-15  & 4.79s    &  27.00    & 1.57e-16   & 0.25 \\
	linearfit               &  2.58e-16  & 7.59s    &  1.08   	& 8.50e-17    & 0.13 \\
	jetApprox\tnote{c}	&  5.72e-15  & 13.10s    &  15.13  	& 2.11e-15   & 0.20 \\
	SqDistPtSeg\tnote{a}	&  4.27e-13  & 15.04s    &  317.58 	& 6.20e-14    & 7.14 \\
	Squareroot\tnote{b}	&  1.22e-15  & 0.52s    &  2.68   	& 7.14e-16   & -- \\
	Styblinski\tnote{c}	&  3.02e-14  & 6.57s    &  51.62  	& 6.03e-15   & 3.21e-07 \\
	Test2\tnote{a}		&  7.90e-14  & 12.71s    &  102.00  	& 3.27-14    & 0.69 \\
	ClosestPoint\tnote{a} 	&  9.98e-16  & 6.30s    &  4.59   	& 5.47e-16   & 2.6e-07 \\
	SpherePt\tnote{a}   	&  6.43e-15  & 66.00s   &  0.87   	& 2.98e-15   & 1.02e-06 \\
	lead-lag\tnote{b}	&  1.66e-16  & 25.90s    &  0.04   	& 7.19e-17   & 1.51e-04 \\
	EigenSphere\tnote{a}  	&  1.67e-15  & 480.00s  &  2.16   	& 5.44e-16   & 1.37e-07 \\
	Jacobi\tnote{a}		&  1.61e-13  & 22.00s   &  81.33  	& 5.89e-14   & -- \\
	Mol-Dyn\tnote{a}	&  2.94e-15  & 138.00s  &  2.11   	& 1.24e-15   & 5.67e-06 \\
	Gram-Schmidt\tnote{b}	&  2.42e-16  & 99.00s   &  0.02 	& 5.28e-17   & 8.37e-04 \\
	Ray Tracing\tnote{a}	&  4.70e-14  & 331.00s  &  87.70   	&  1.15e-14  & 2.90e-08 \\
	ExtremePoint\tnote{a}	&  1.69e-14  & 30.00s   &  115.50  	&  1.21e-14  & 4.52e-06 \\  
	SmartRoot\tnote{b}	&  {\bf inf} & --       &  {\bf inf}   	& 3.84e-16   & 1.26e-07 \\
	Poisson\tnote{a}	&  1.0e-17   & 120.00s  &  0.97   	& 4.52e-18    & -- \\
	\end{tabular}
	\begin{tablenotes}
		\item[a] New benchmarks introduced for conditional codes
		\item[b] Benchmarks ported from FPBench
		\item[c] Benchmarks ported from PRECiSA
	\end{tablenotes}
	\caption{\label{tab:seesaw-solver-compare} \satire evaluation for worst case error bounds and
    instability jumps
    for code with conditionals. }
	\end{threeparttable}
	\end{table*}
\endgroup

\subsubsection{\textbf{Empirical testing and statistical profiling}}
%

To empirically validate the tightness of our error estimates, we cross-checked them statistically by sampling the error function over a population of intervals. This approach offers an alternative perspective on the error, free from potential over-approximations that solvers might introduce. \satire's bounds were consistently conservative and, in many cases, within the same order of magnitude as the observed empirical bounds. On average, the bounds deviated by no more than two orders of magnitude, except in specific instances where a lack of awareness of domain constraints could result in extreme values (e.g., \texttt{inf}).

Instead of relying solely on global optimization of symbolic error expressions, one can also evaluate them through statistical sampling.
This method provides a practical view of the underlying error expression without the over-approximations often introduced by optimization techniques. A significant discrepancy between the rigorous worst-case error bounds and the statistically sampled bounds can offer valuable insights, particularly when worst-case errors are driven by a few outlier values. The bounds obtained through sampling error expressions should lie between the rigorous worst-case bounds and the exact runtime error observed on concrete data points. This relationship exists because error expressions themselves represent over-approximations of actual runtime rounding errors.

In Table~\ref{tab:seesaw-solver-compare}, the column labeled `Prof Error Expr' summarizes the error estimates derived from sampling the rigorous error expression bounds.
These sampling methods are often preferred over shadow-value-based error estimation, which requires duplicating the code at two different precision levels—a process that is not always feasible.
Our sampling approach employs a uniform sampling strategy.
Our empirical profiling provides reassurance that \satire is sound in the sense that it does not fall below the empirical bound but also stays close, showing the tightness of its rigorous bounds.

\subsubsection{\bf Instability analysis}

We report the maximum potential output deviations caused by instability called an instability jump. \satire's Conditional Computation Graph generation facilitates the extraction of guarded path constraints that identify valid program paths. However, these path conditions often contain numerous atomic inequalities involving larger floating-point expressions. We rigorously analyze each Boolean predicate within the path conditions and systematically weaken individual constraints to account for the instability region (the "gray zone") caused by round-off errors. Let $P$ represent the original atomic inequality and $P^\prime$ denote the weakened constraint obtained by incorporating rounding errors into $P$. Under this transformation, the expression $(P^\prime \wedge \lnot P^\prime)$ does not evaluate to `False.’ Instead, it explicitly identifies the instability region where discrepancies between real-number execution and floating-point execution may arise, potentially leading to instability.

%
%
For a Boolean predicate composed of multiple inequalities, determining the combined cover for the instability region is generally a non-trivial task. However, by applying our predicate weakening approach to each atomic formula individually, this process can be seamlessly integrated into the overall domain constraints in a general manner. We compute the function difference along pairs of program paths within these instability regions and report the maximum value for instability quantification. It is crucial that the instability expression is solved only within the bounds of the identified instability constraints. Otherwise, it would yield overly inflated results across the entire interval, as illustrated in Table~\ref{tab:seesaw-solver-compare}.

We also performed {\em empirical testing} to explore the input domain over randomly sampled  points to hit regions that trigger these instabilities.
In Table~\ref{tab:seesaw-solver-compare}, the last column lists the maximum instability jump observed empirically using 10M tests per benchmark. The `-' entries did not hit any instability issues in our empirical tests.





\section{Mixed-Precision Evaluation}
\label{sec:mix-prec-eval}

\begin{table}
    \centering
    \begin{tabular}{crrrrrrrrr}
    \toprule
         \multirow{2}{*}{Benchmarks} & \multirow{2}{*}{\shortstack[1]{Num \\OPs \\ (in \\1000's)}} & \multirow{2}{*}{\shortstack[1]{\% Single \\Precision \\Operations}} & \multicolumn{2}{c}{Uniform Single Precision} & \multicolumn{2}{c}{Mixed Precision} & \multirow{2}{*}{\shortstack[1]{Empirical \\ Error bounds for \\ Mixed Precision}} \\

  \cmidrule(lr){4-5}  \cmidrule(lr){6-7}

  & & & \shortstack[1]{\textbf{Absolute} \\{\textbf{Error Bound}}} & \shortstack[1]
  {Execution \\Time\\ (secs)} & \shortstack[1]{Absolute \\{Error Bound}} & \shortstack[1]{Execution \\Time\\(secs)} & \\
    \midrule
         Scan(1024pt) & 2 & 50 & 1.00e-03 & \textbf{179.14} & 8.70e-04 & 180.04 & 5.12e-04 \\
         Scan(4096pt) & 8 & 50 & 5.00e-02 & 4021.24 & 3.17e-02 & \textbf{3375.79} & 6.93e-03 \\
         FFT-1024pt & 81 & 20 & 9.03e-04 & \textbf{213.96} & 3.42e-04 & 217.82 & 1.84e-04 \\
         FFT-4096pt & 393 & 20 & 7.92e-03 & \textbf{3850.17} & 1.51e-03 & 3910.74 & 9.35e-04 \\
         Advection & 453 & 50 & 6.41e-05 & 16454.75 & 3.17e-05 & \textbf{16342.72} & 4.41e-06 \\
         FDTD & 192 & 50 & 3.38e-04 & \textbf{10294.95} & 3.18e-04 & 13473.03 & 1.01e-04 \\
         matmul-64 & 520 & 10 & 8.10e-04 & \textbf{35.18} & 5.10e-04 & 38.29 & 2.73e-05 \\
         matmul-128 & 4177 & 10 & 6.62e-03 & \textbf{214.22} & 1.99e-03 & 218.54 & 8.25e-04 \\
         Serial Sum & 1 & 50 & 2.73e-05 & \textbf{4.92} & 1.45e-05 & 5.71 & 7.68e-06 \\
         Reduction & 1 & 50 & 6.10e-04 & \textbf{60.81} & 6.10e-04 & 62.77 & 9.51e-05 \\
         Poly-50 & 1.3 & 50 & 1.50e-05 & \textbf{4.36} & 1.45e-05 & 5.72 & 8.27e-06 \\
         Horner-50 & 0.1 & 50 & 8.70e-05 & \textbf{3.56} & 3.93e-05 & 5.11 & 5.12e-06 \\
    \end{tabular}
    \caption{\satire evaluated worst case absolute error bounds on larger uniform single-precision and mixed-precision benchmarks}
    \label{tab:error-bound-comparison}
\end{table}

We evaluate benchmarks from \satire's test suite which cover stencils (e.g., Finite-Difference Time-Domain FDTD), iterative solvers, Fast Fourier Transforms (FFT), and matrix multiplication. We report and compare error bounds and performance of uniform single precision workloads with corresponding mixed precision (single-precision and double-precision) workloads created by increasing precision of some randomly chosen expressions.

\subsection{\bf Error Bound, Performance Comparison and Empirical Testing}
Table~\ref{tab:error-bound-comparison} summarizes the error bounds obtained for the larger benchmarks without any abstractions. The ``Num OPs'' column indicates the number of operators in a single expression tree created by unrolling loops to varying sizes in these benchmarks. The ``\% Single Precision Operations'' column specifies the proportion of single-precision operations relative to the total number of operations. Across all benchmarks, mixed precision consistently achieves lower error bounds compared to uniform single precision, highlighting its effectiveness in reducing numerical errors while maintaining computational efficiency.

Stencil benchmarks exhibit particularly tight bounds, with errors on the order of $e-04$, aligning closely with empirical results. In contrast, prefix-sum (Scan) benchmarks show empirical error bounds that are one order of magnitude tighter than the calculated worst-case bounds.
Benchmarks with a larger number of operations, such as FFT-4096pt and matmul-128, demonstrate significant improvements in precision with mixed precision, further emphasizing its scalability and accuracy benefits. For smaller benchmarks like "Serial Sum" and "Reduction," the improvements with mixed precision are moderate. However, its relative impact becomes considerably more pronounced for computationally intensive benchmarks like matrix multiplication and stencils, where the potential for error accumulation is higher.


For the majority of benchmarks, the execution times for mixed precision and uniform single precision are comparable. This indicates that mixed precision achieves improved error bounds without incurring significant computational overhead.
For computationally intensive benchmarks such as "Advection" and "FDTD," execution times are higher for both precision levels. However, the performance gap between single and mixed precision remains minimal, highlighting the efficiency of mixed precision even for workloads with high computational demands.
Benchmarks with smaller workloads, such as "Serial Sum" and "Poly-50," exhibit negligible differences in execution times between the two precision configurations. Meanwhile, larger benchmarks continue to demonstrate computational efficiency with mixed precision, further underscoring its suitability for complex and large-scale workloads.


We performed empirical testing to assess the tightness and empirical soundness of \satire's worst-case absolute error bounds for uniform single-precision and mixed-precision benchmarks. The oracle used for validation was GCC's quad precision operations.
Using random sampling, $10^6$ inputs were generated across specified intervals and tested against these benchmarks. \satire's error bounds consistently demonstrated conservativeness, with average deviations from the empirical bounds being no greater than one order of magnitude.
These results validate \satire's ability to provide reliable error estimates, ensuring robust numerical analysis while maintaining conservative yet realistic bounds across both precision configurations.

\subsection{\bf Iterative Method - Conjugate Gradient}

The Conjugate Gradient (CG) method is an iterative optimization algorithm widely used to solve systems of linear equations, particularly those of the form $Ax=b$. It plays a significant role in numerical linear algebra and optimization. The steps of the CG method are outlined below:

\begin{algorithm}[H]
\caption{Conjugate Gradient Method}
\begin{multicols}{2}
\begin{algorithmic}[1]
\STATE Compute \(r_0 = b - Ax_0\)
\STATE Set \(p_0 = r_0\)
\STATE Set \(k = 0\)
\WHILE{Convergence condition not met}
    \STATE Compute step size: \(\alpha_k = \frac{r_k^T r_k}{p_k^T A p_k}\)
    \STATE Update solution: \(x_{k+1} = x_k + \alpha_k p_k\)
    \STATE Update residual: \(r_{k+1} = r_k - \alpha_k A p_k\)
    \IF{Convergence condition met}
        \STATE \textbf{break}
    \ENDIF
    \STATE Compute direction scaling: \(\beta_k = \frac{r_{k+1}^T r_{k+1}}{r_k^T r_k}\)
    \STATE Update search direction: \(p_{k+1} = r_{k+1} + \beta_k p_k\)
    \STATE Increment \(k = k + 1\)
\ENDWHILE
\RETURN \(x_k\) (approximate solution)
\end{algorithmic}
\end{multicols}
\end{algorithm}

\begin{table}
    \centering
    \begin{tabular}{rrrrrrrrr}
    \toprule
         \multirow{3}{*}{Iteration} & \multicolumn{4}{c}{Alpha} & \multicolumn{4}{c}{Beta} \\

  \cmidrule(lr){2-5}  \cmidrule(lr){6-9}
  & \multicolumn{2}{c}{Single Precision} & \multicolumn{2}{c}{\textbf{Mixed Precision}} & \multicolumn{2}{c}{Single Precision} & \multicolumn{2}{c}{\textbf{Mixed Precision}} \\
  \cmidrule(lr){2-3} \cmidrule(lr){4-5} \cmidrule(lr){6-7} \cmidrule(lr){8-9}
  & Abs Err & Rel Err & Abs Err & Rel Err & Abs Err & Rel Err & Abs Err & Rel Err\\
    \midrule
         1 & 1.27e-01 & 3.64e-01 & 1.24e-01 & 2.89e-01 & 2.10e-01 & 1.08e-00 & 2.03e-02 & 4.25e-01 \\
         2 & 8.66e-02 & 2.97e-01 & 7.92e-02 & 2.79e-01 & 6.29e-01 & 1.70e-00 & 6.46e-02 & 1.83e-01 \\
         3 & 2.88e-02 & 4.29e-00 & 2.56e-02 & 3.81e-00 & 7.20e-01 & 4.92e-00 & 5.40e-01 & 1.64e-00 \\
         4 & 3.51e-02 & 4.74e-00 & 3.11e-02 & 4.20e-00 & 9.10e-01 & 2.48e-00 & 7.60e-01 & 1.47e-00 \\
         5 & 4.72e-02 & 5.49e-00 & 4.19e-02 & 4.87e-00 & 9.10e-01 & 2.55e-00 & 7.50e-01 & 1.42e-00 \\
         6 & 6.85e-02 & 6.48e-00 & 6.08e-02 & 5.75e-00 & 7.70e-01 & 8.50e-00 & 4.90e-01 & 5.30e-00 \\
         7 & 1.06e-01 & 7.69e-00 & 9.43e-02 & 6.84e-00 & 9.50e-01 & 1.70e-00 & 6.80e-01 & 2.41e-00 \\
         8 & 1.71e-01 & 8.96e-00 & 1.52e-01 & 7.96e-00 & 8.40e-01 & 3.30e-00 & 6.50e-01 & 1.47e-00 \\
         9 & 2.65e-01 & 9.68e-00 & 2.36e-01 & 8.62e-00 & 1.13e-00 & 8.57e-00 & 5.76e-01 & 1.36e-00 \\
         10 & 3.15e-01 & 8.33e-00 & 2.82e-01 & 7.46e-00 & 1.89e-00 & 2.86e-00 & 1.66e-00 & 1.86e-00 \\
         11 & 1.56e-01 & 3.61e-00 & 1.42e-01 & 3.29e-00 & 1.91e-00 & 2.10e-00 & 1.86e-00 & 2.15e-00 \\
         12 & 1.60e-01 & 4.39e-00 & 1.43e-01 & 3.92e-00 & 1.80e-00 & 2.61e-00 & 1.68e-00 & 1.68e-00 \\
         13 & 1.82e-01 & 7.34e-00 & 1.62e-01 & 6.53e-00 & 1.96e-00 & 2.08e-00 & 1.94e-00 & 1.94e-00 \\
         14 & 1.89e-01 & 1.17e-01 & 1.66e-01 & 1.03e-01 & 2.13e-00 & 2.37e-00 & 1.89e-00 & 1.89e-00 \\
         15 & 1.93e-01 & 1.76e-01 & 1.73e-01 & 1.57e-01 & 1.99e-00 & 2.27e-00 & 1.87e-00 & 1.87e-00 \\
         16 & 2.54e-01 & 3.19e-01 & 2.38e-01 & 2.99e-01 & 2.00e-00 & 3.66e-00 & 1.54e-00 & 1.54e-00 \\
         17 & 2.98e-01 & 4.88e-01 & 2.86e-01 & 4.68e-01 & 1.99e-00 & 4.57e-00 & 1.43e-00 & 1.43e-00 \\
         18 & 3.42e-01 & 6.95e-01 & 3.32e-01 & 6.75e-01 & 1.98e-00 & 1.98e-00 & 1.95e-00 & 1.95e-00 \\
         19 & 4.18e-01 & 1.01e-02 & 3.97e-01 & 9.62e-01 & 1.96e-00 & 6.75e-00 & 1.29e-00 & 1.29e-00 \\
         20 & 5.00e-01 & 1.40e-02 & 4.98e-01 & 1.39e-02 & 1.69e-00 & 2.90e-00 & 1.58e-00 & 2.27e-00 \\
    \end{tabular}
    \caption{Absolute and Relative Error bounds of $\alpha$ and $\beta$ parameters of a uniform single-precision and mixed-precision conjugate gradient program}
    \label{tab:iterative-refinement}
\end{table}


We examine the effects of mixed precision on the error bounds of key components in the CG method. Specifically, we apply \satire to a conjugate gradient benchmark running on a 27-dimensional linear system over 20 iterations. Two configurations are compared: a uniform single precision implementation and a mixed precision implementation where double precision is used for the residual $r_{k+1}$ and solution $x_{k+1}$ computations at each iteration. These double-precision components account for 16.95\% of a total of 21,504 operations. Although \satire does not directly provide the capability to estimate relative errors, we 
compute an approximation
by taking
the ratio of the absolute error bound and the smallest absolute value of the function output in question. This provides a conservative approximation of the worst case relative error bound.

Table~\ref{tab:iterative-refinement} presents the error bounds for Alpha (step size) and Beta (direction scaling). The results show that the error bounds for both Alpha and Beta generally trend upward with increasing iterations, reflecting the cumulative nature of error propagation in iterative methods. 
The mixed precision implementation consistently achieves lower error bounds than the single precision version across all iterations. This demonstrates the effectiveness of mixed precision in improving numerical accuracy while containing error growth. 
Beta computations exhibit larger relative errors compared to Alpha, particularly in single precision. This highlights the sensitivity of Beta, which directly influences the search direction, to precision levels. Mixed precision mitigates this sensitivity effectively, keeping relative errors within acceptable bounds.
Mixed precision not only reduces the absolute and relative errors but also moderates the erratic growth observed in single precision. This makes mixed precision particularly beneficial for iterative methods like the CG method, where cumulative error growth can impact convergence.

\section{Other Related Work}
\label{sec:conclusion}

Among existing methods, Rosa~\cite{Darulova2017} and FPTaylor~\cite{fptaylor} are the closest to our work, as both extend Taylor forms~\cite{taylorforms} to 
enable rigorous floating-point error estimation. 
Rosa propagates errors in a numeric affine form and employs SMT solvers to obtain tight bounds. 
FPTaylor constructs full symbolic Taylor forms, which are then optimized globally.
\satire outperforms these tools in bound-tightness and scalability.

Many tools (e.g.,  \fptaylor\cite{fptaylor}, \satire~\cite{2020_Das},
 \precisa~\cite{precisa})
in this area use a global optimizer to estimate the upper bound
of roundoff error.
These optimizers work over rectangular input domains
using variants of branch-and-bound~\cite{alliot2012finding,fptaylor,gelpia-github}. \seesaw~\cite{2021_Das} presents comparative results 
for \seesaw, \precisa and \fluctuat for benchmarks containing
straight-line codes with conditions.
Comparison is presented for 
the total execution time, 
absolute error bounds and 
the bounds obtained on the output instability.

Inferring errors across loops is fraught with difficulties---the main one being the inability to infer loop invariants that are true under the floating-point semantics.
Darulova et al.~\cite{eva-loop-invariant} propose handling loops by synthesizing loop invariants in the presence of round-off errors, but their method is limited in scalability and applicable to ellipsoid domains only. In \satire, although we do need to unroll loops, we are not limited to small examples or to ellipsoid domains and can handle linear as well as non-linear applications of a wide variety.

The $\Lambda_{num}$ system developed by Kellison et al.~\cite{2025_ariel} performs error analysis using a type-based approach, disregards underflow and overflow and works only on positive numbers.
\fpguard~\cite{2025_fpguard}, an extension of \satire, attempts to solve the problem of singularities in its domain for examples like x/(x-y) and can automatically exclude problematic domain points to avoid degeneracy.

Our work is the first we know that employs the idea of paving to
deal with non-rectangular domains and still be able to use a backend
rectangular domain optimizer.
In terms of solver technologies,
Rosa~\cite{rosa} propagates 
errors in numeric affine form and uses SMT solvers
to obtain tight bounds. The idea of weakening predicates is mentioned in previous work
(e.g., Darulova~\cite{darulova-phd}).
However, in our work, we provide the first implementation (we know)
of these ideas within a symbolic reverse-mode AD framework.


Automatic differentiation (AD)\cite{automatic-diff-hovland} is a widely used method in scientific computing for derivative evaluation, 
relying on chain-rule-based techniques. ADAPT\cite{adapt} presents a scalable approach for mixed-precision tuning in HPC applications but is limited to concrete data points. 
It utilizes Codipack~\cite{codipack} to perform reverse-mode AD and obtain derivative values. In contrast, \satire implements its own symbolic library for reverse-mode AD, supporting analysis over input interval ranges to estimate output errors across the entire input space rigorously. 
This provides a mechanism for formal specification inference, aiding future code users in reliably adapting software to new environments and precision regimes—an increasingly important requirement in HPC.

Although \satire does not perform precision tuning itself, it offers new capabilities that can complement existing tuning tools such as \precimonious~\cite{sc13-precimonious}, FPTuner~\cite{fptuner}, POP~\cite{2021_dorra_pop}, and ADAPT. 
By integrating \satire, these tools can refine code accuracy in a specification-driven manner by focusing on specific code regions and interval sub-ranges. 
Furthermore, \satire can enhance tools such as Herbie~\cite{herbie} by guiding them to rewrite subexpressions in a more goal-directed manner, improving numerical precision where needed. Similarly, the methods described in~\cite{DBLP:journals/sttt/DamoucheMC17} can also benefit from \satire's capabilities.


\subsection{Concluding Remarks}

We presented \satire---a tool
that combines rigorous absolute error estimation aided by abstraction and incorporates
methods to handle conditional
cexpressions as well as
mixed-precision allocation.
We showcased \satire on the
design of a Conjugate Gradient
Iterative Solver. 
Our results are indicative of
a tool that is already useful for designers who care about developing numerical code informed by rigorous error estimates.

A direct formulation of relative error estimation is presented in~\cite{fptaylor}; the main difficulty one faces is that the error expressions now involve a ratio (division by the true value in symbolic form).
We have faced difficulties in solving such expressions using available optimizers.
We find the method used here---doing the ratioing after the fact to be a reasonable practical alternative that, while not sound, is sufficiently indicative, we believe.

We did not treat subnormals or floating-point exceptions in this work.
However, our group is active in exploring this topic in the context of GPU computations where the hardware does not trap exceptions---hence requiring very efficiently engineered binary instrumentation methods~\cite{gpu-fpx}.

A tool such as \satire takes up the ambitious problem of evaluating the error across multple input intervals.
The fate of this depends on how well the optimizer performs---and whether the optimizer itself quits.
In~\cite{2025_fpguard}, we address this issue by ruling out input intervals where the optimizer runs into undefined cases such as division by zero.
A more interesting and practical solution might be to run a concrete automatic differentiation tool such as Enzyme~\cite{2021_enzyme} at these domain points.
In fact, tools such as Enzyme might also help the user achieve even higher scale by mixing symbolic and concrete automatic differentiation.

One of the bottlenecks that remains is the performance of the optimizer.
One intriguing possibility is to run  multiple mixed-precision assignments and
multiple abstraction selections
in parallel, selecting those that return quickly.
This might be especially be handy in a design situation where rapid error analysis might enable a designer to proceed binding decisions early---and avoid expressions that prove to be hard to optimize.
These directions form exciting avenues for our future work.



\bibliographystyle{ACM-Reference-Format}
\bibliography{references}


\begin{thebibliography}{37}


\ifx \showCODEN    \undefined \def \showCODEN     #1{\unskip}     \fi
\ifx \showDOI      \undefined \def \showDOI       #1{#1}\fi
\ifx \showISBNx    \undefined \def \showISBNx     #1{\unskip}     \fi
\ifx \showISBNxiii \undefined \def \showISBNxiii  #1{\unskip}     \fi
\ifx \showISSN     \undefined \def \showISSN      #1{\unskip}     \fi
\ifx \showLCCN     \undefined \def \showLCCN      #1{\unskip}     \fi
\ifx \shownote     \undefined \def \shownote      #1{#1}          \fi
\ifx \showarticletitle \undefined \def \showarticletitle #1{#1}   \fi
\ifx \showURL      \undefined \def \showURL       {\relax}        \fi
\providecommand\bibfield[2]{#2}
\providecommand\bibinfo[2]{#2}
\providecommand\natexlab[1]{#1}
\providecommand\showeprint[2][]{arXiv:#2}

\bibitem[Alliot et~al\mbox{.}(2012)]%
        {alliot2012finding}
\bibfield{author}{\bibinfo{person}{Jean-Marc Alliot}, \bibinfo{person}{Nicolas Durand}, \bibinfo{person}{David Gianazza}, {and} \bibinfo{person}{Jean-Baptiste Gotteland}.} \bibinfo{year}{2012}\natexlab{}.
\newblock \showarticletitle{Finding and Proving the Optimum: Cooperative Stochastic and Deterministic Search}. In \bibinfo{booktitle}{\emph{Proceedings of the 20th European Conference on Artificial Intelligence (ECAI)}}. \bibinfo{publisher}{ACM}, \bibinfo{pages}{55--60}.
\newblock
\urldef\tempurl%
\url{https://doi.org/10.3233/978-1-61499-098-7-55}
\showDOI{\tempurl}


\bibitem[Ben~Khalifa and Martel(2021)]%
        {2021_dorra_pop}
\bibfield{author}{\bibinfo{person}{Dorra Ben~Khalifa} {and} \bibinfo{person}{Matthieu Martel}.} \bibinfo{year}{2021}\natexlab{}.
\newblock \showarticletitle{An Evaluation of POP Performance for Tuning Numerical Programs in Floating-Point Arithmetic}. In \bibinfo{booktitle}{\emph{2021 4th International Conference on Information and Computer Technologies (ICICT)}}. \bibinfo{pages}{69--78}.
\newblock
\urldef\tempurl%
\url{https://doi.org/10.1109/ICICT52872.2021.00019}
\showDOI{\tempurl}


\bibitem[Berg et~al\mbox{.}(2008)]%
        {compgeo}
\bibfield{author}{\bibinfo{person}{Mark~de Berg}, \bibinfo{person}{Otfried Cheong}, \bibinfo{person}{Marc~van Kreveld}, {and} \bibinfo{person}{Mark Overmars}.} \bibinfo{year}{2008}\natexlab{}.
\newblock \bibinfo{booktitle}{\emph{Computational Geometry: Algorithms and Applications} (\bibinfo{edition}{3rd ed.} ed.)}.
\newblock \bibinfo{publisher}{Springer-Verlag TELOS}, \bibinfo{address}{Santa Clara, CA, USA}.
\newblock
\showISBNx{3540779736}


\bibitem[Bischof et~al\mbox{.}(2008)]%
        {automatic-diff-hovland}
\bibfield{editor}{\bibinfo{person}{{C.H.} Bischof}, \bibinfo{person}{{H.M.} Buker}, \bibinfo{person}{Paul Hovland}, \bibinfo{person}{{U.} Naumann}, {and} \bibinfo{person}{{J.} Utke}} (Eds.). \bibinfo{year}{2008}\natexlab{}.
\newblock \bibinfo{booktitle}{\emph{Advances in Automatic Differentiation}}.
\newblock \bibinfo{publisher}{Springer}.
\newblock
\newblock
\shownote{ISBN : 978-3-540-68935-5}.


\bibitem[Chiang et~al\mbox{.}(2017)]%
        {fptuner}
\bibfield{author}{\bibinfo{person}{Wei-Fan Chiang}, \bibinfo{person}{Mark Baranowski}, \bibinfo{person}{Ian Briggs}, \bibinfo{person}{Alexey Solovyev}, \bibinfo{person}{Ganesh Gopalakrishnan}, {and} \bibinfo{person}{Zvonimir Rakamariundefined}.} \bibinfo{year}{2017}\natexlab{}.
\newblock \showarticletitle{Rigorous Floating-Point Mixed-Precision Tuning}.
\newblock \bibinfo{journal}{\emph{SIGPLAN Not.}} \bibinfo{volume}{52}, \bibinfo{number}{1} (\bibinfo{date}{Jan.} \bibinfo{year}{2017}), \bibinfo{pages}{300–315}.
\newblock
\showISSN{0362-1340}
\urldef\tempurl%
\url{https://doi.org/10.1145/3093333.3009846}
\showDOI{\tempurl}


\bibitem[Damouche et~al\mbox{.}(2017)]%
        {DBLP:journals/sttt/DamoucheMC17}
\bibfield{author}{\bibinfo{person}{Nasrine Damouche}, \bibinfo{person}{Matthieu Martel}, {and} \bibinfo{person}{Alexandre Chapoutot}.} \bibinfo{year}{2017}\natexlab{}.
\newblock \showarticletitle{Improving the Numerical Accuracy of Programs By Automatic Transformation}.
\newblock \bibinfo{journal}{\emph{International Journal on Software Tools for Technology Transfer (STTT)}} \bibinfo{volume}{19}, \bibinfo{number}{4} (\bibinfo{year}{2017}), \bibinfo{pages}{427--448}.
\newblock
\urldef\tempurl%
\url{https://doi.org/10.1007/s10009-016-0435-0}
\showDOI{\tempurl}


\bibitem[Darulova and Kuncak(2014)]%
        {rosa}
\bibfield{author}{\bibinfo{person}{Eva Darulova} {and} \bibinfo{person}{Viktor Kuncak}.} \bibinfo{year}{2014}\natexlab{}.
\newblock \showarticletitle{Sound Compilation of Reals}. In \bibinfo{booktitle}{\emph{Proceedings of the 41st ACM SIGPLAN-SIGACT Symposium on Principles of Programming Languages (POPL)}} (San Diego, California, USA). \bibinfo{publisher}{ACM}, \bibinfo{pages}{235--248}.
\newblock


\bibitem[Darulova and Kuncak(2017)]%
        {Darulova2017}
\bibfield{author}{\bibinfo{person}{Eva Darulova} {and} \bibinfo{person}{Viktor Kuncak}.} \bibinfo{year}{2017}\natexlab{}.
\newblock \showarticletitle{Towards a Compiler for Reals}.
\newblock \bibinfo{journal}{\emph{ACM Trans. Program. Lang. Syst.}} \bibinfo{volume}{39}, \bibinfo{number}{2} (\bibinfo{year}{2017}), \bibinfo{pages}{8:1--8:28}.
\newblock
\urldef\tempurl%
\url{https://doi.org/10.1145/3014426}
\showDOI{\tempurl}


\bibitem[Das et~al\mbox{.}(2020a)]%
        {2020_Das}
\bibfield{author}{\bibinfo{person}{Arnab Das}, \bibinfo{person}{Ian Briggs}, \bibinfo{person}{Ganesh Gopalakrishnan}, \bibinfo{person}{Sriram Krishnamoorthy}, {and} \bibinfo{person}{Pavel Panchekha}.} \bibinfo{year}{2020}\natexlab{a}.
\newblock \showarticletitle{Scalable yet Rigorous Floating-Point Error Analysis}. In \bibinfo{booktitle}{\emph{SC20: International Conference for High Performance Computing, Networking, Storage and Analysis}}. \bibinfo{pages}{1--14}.
\newblock
\urldef\tempurl%
\url{https://doi.org/10.1109/SC41405.2020.00055}
\showDOI{\tempurl}


\bibitem[Das et~al\mbox{.}(2020b)]%
        {fpd}
\bibfield{author}{\bibinfo{person}{Arnab Das}, \bibinfo{person}{Sriram Krishnamoorthy}, \bibinfo{person}{Ian Briggs}, \bibinfo{person}{Ganesh Gopalakrishnan}, {and} \bibinfo{person}{Ramakrishna Tipireddy}.} \bibinfo{year}{2020}\natexlab{b}.
\newblock \bibinfo{title}{Efficient Reasoning About Stencil Programs Using Selective Direct Evaluation}.
\newblock
\newblock
\showeprint[arxiv]{2004.04359}~[cs.DC]


\bibitem[Das et~al\mbox{.}(2020c)]%
        {fpdetect}
\bibfield{author}{\bibinfo{person}{Arnab Das}, \bibinfo{person}{Sriram Krishnamoorthy}, \bibinfo{person}{Ian Briggs}, \bibinfo{person}{Ganesh Gopalakrishnan}, {and} \bibinfo{person}{Ramakrishna Tipireddy}.} \bibinfo{year}{2020}\natexlab{c}.
\newblock \showarticletitle{FPDetect: Efficient Reasoning About Stencil Programs Using Selective Direct Evaluation}.
\newblock \bibinfo{journal}{\emph{{ACM} Trans. Archit. Code Optim.}} \bibinfo{volume}{17}, \bibinfo{number}{3} (\bibinfo{year}{2020}), \bibinfo{pages}{19:1--19:27}.
\newblock
\urldef\tempurl%
\url{https://dl.acm.org/doi/10.1145/3402451}
\showURL{%
\tempurl}


\bibitem[Das et~al\mbox{.}(2021)]%
        {2021_Das}
\bibfield{author}{\bibinfo{person}{Arnab Das}, \bibinfo{person}{Tanmay Tirpankar}, \bibinfo{person}{Ganesh Gopalakrishnan}, {and} \bibinfo{person}{Sriram Krishnamoorthy}.} \bibinfo{year}{2021}\natexlab{}.
\newblock \showarticletitle{Robustness Analysis of Loop-Free Floating-Point Programs via Symbolic Automatic Differentiation}. In \bibinfo{booktitle}{\emph{2021 IEEE International Conference on Cluster Computing (CLUSTER)}}. \bibinfo{pages}{481--491}.
\newblock
\urldef\tempurl%
\url{https://doi.org/10.1109/Cluster48925.2021.00055}
\showDOI{\tempurl}


\bibitem[Davis and Hu(2011)]%
        {ufl}
\bibfield{author}{\bibinfo{person}{Timothy~A. Davis} {and} \bibinfo{person}{Yifan Hu}.} \bibinfo{year}{2011}\natexlab{}.
\newblock \showarticletitle{The University of Florida Sparse Matrix Collection}.
\newblock \bibinfo{journal}{\emph{ACM Trans. Math. Softw.}} \bibinfo{volume}{38}, \bibinfo{number}{1}, Article \bibinfo{articleno}{1} (\bibinfo{date}{Dec.} \bibinfo{year}{2011}), \bibinfo{numpages}{25}~pages.
\newblock
\showISSN{0098-3500}
\urldef\tempurl%
\url{https://doi.org/10.1145/2049662.2049663}
\showDOI{\tempurl}


\bibitem[Ericson(2004)]%
        {realtime}
\bibfield{author}{\bibinfo{person}{Christer Ericson}.} \bibinfo{year}{2004}\natexlab{}.
\newblock \bibinfo{booktitle}{\emph{Real-Time Collision Detection}}.
\newblock \bibinfo{publisher}{CRC Press, Inc.}, \bibinfo{address}{USA}.
\newblock
\showISBNx{1558607323}


\bibitem[Gelpia(2017)]%
        {gelpia-github}
Gelpia \bibinfo{year}{2017}\natexlab{}.
\newblock \bibinfo{title}{Gelpia: A Global Optimizer for Real Functions}.
\newblock
\newblock
\urldef\tempurl%
\url{https://github.com/soarlab/gelpia}
\showURL{%
Retrieved October 13, 2017 from \tempurl}


\bibitem[Goldberg(1991)]%
        {Goldberg1991}
\bibfield{author}{\bibinfo{person}{David Goldberg}.} \bibinfo{year}{1991}\natexlab{}.
\newblock \showarticletitle{What Every Computer Scientist Should Know About Floating-Point Arithmetic}.
\newblock \bibinfo{journal}{\emph{Comput. Surveys}} \bibinfo{volume}{23}, \bibinfo{number}{1} (\bibinfo{date}{March} \bibinfo{year}{1991}), \bibinfo{pages}{5--48}.
\newblock
\showISSN{0360-0300}
\urldef\tempurl%
\url{https://doi.org/10.1145/103162.103163}
\showDOI{\tempurl}


\bibitem[Goualard(2017)]%
        {gaol}
\bibfield{author}{\bibinfo{person}{Fr\'{e}d\'{e}ric Goualard}.} \bibinfo{year}{2017}\natexlab{}.
\newblock \bibinfo{booktitle}{\emph{GAOL (Not Just Another Interval Library)}}.
\newblock
\urldef\tempurl%
\url{http://frederic.goualard.net/\#research-software}
\showURL{%
Retrieved October 13, 2017 from \tempurl}


\bibitem[Higham({[n.\,d.]})]%
        {higham}
\bibfield{author}{\bibinfo{person}{Nick Higham}.} \bibinfo{year}{[n.\,d.]}\natexlab{}.
\newblock \bibinfo{booktitle}{\emph{Accuracy and Stability of Numerical Algorithms, 2nd Edition}}.
\newblock \bibinfo{publisher}{SIAM}.
\newblock


\bibitem[Izycheva et~al\mbox{.}(2020)]%
        {eva-loop-invariant}
\bibfield{author}{\bibinfo{person}{Anastasiia Izycheva}, \bibinfo{person}{Eva Darulova}, {and} \bibinfo{person}{Helmut Seidl}.} \bibinfo{year}{2020}\natexlab{}.
\newblock \showarticletitle{Counterexample- and Simulation-Guided Floating-Point Loop Invariant Synthesis}. In \bibinfo{booktitle}{\emph{Static Analysis}}, \bibfield{editor}{\bibinfo{person}{David Pichardie} {and} \bibinfo{person}{Mihaela Sighireanu}} (Eds.). \bibinfo{publisher}{Springer International Publishing}, \bibinfo{address}{Cham}, \bibinfo{pages}{156--177}.
\newblock
\showISBNx{978-3-030-65474-0}


\bibitem[Kellison and Hsu(2024)]%
        {2025_ariel}
\bibfield{author}{\bibinfo{person}{Ariel~E. Kellison} {and} \bibinfo{person}{Justin Hsu}.} \bibinfo{year}{2024}\natexlab{}.
\newblock \showarticletitle{Numerical Fuzz: A Type System for Rounding Error Analysis}.
\newblock \bibinfo{journal}{\emph{Proc. ACM Program. Lang.}} \bibinfo{volume}{8}, \bibinfo{number}{PLDI}, Article \bibinfo{articleno}{226} (\bibinfo{date}{June} \bibinfo{year}{2024}), \bibinfo{numpages}{25}~pages.
\newblock
\urldef\tempurl%
\url{https://doi.org/10.1145/3656456}
\showDOI{\tempurl}


\bibitem[Lee et~al\mbox{.}(2018)]%
        {aiken-math-dot-h}
\bibfield{author}{\bibinfo{person}{Wonyeol Lee}, \bibinfo{person}{Rahul Sharma}, {and} \bibinfo{person}{Alex Aiken}.} \bibinfo{year}{2018}\natexlab{}.
\newblock \showarticletitle{On automatically proving the correctness of math.h implementations}.
\newblock \bibinfo{journal}{\emph{{PACMPL}}} \bibinfo{volume}{2}, \bibinfo{number}{{POPL}} (\bibinfo{year}{2018}), \bibinfo{pages}{47:1--47:32}.
\newblock
\urldef\tempurl%
\url{https://doi.org/10.1145/3158135}
\showDOI{\tempurl}


\bibitem[Li et~al\mbox{.}(2023)]%
        {gpu-fpx}
\bibfield{author}{\bibinfo{person}{Xinyi Li}, \bibinfo{person}{Ignacio Laguna}, \bibinfo{person}{Bo Fang}, \bibinfo{person}{Katarzyna Swirydowicz}, \bibinfo{person}{Ang Li}, {and} \bibinfo{person}{Ganesh Gopalakrishnan}.} \bibinfo{year}{2023}\natexlab{}.
\newblock \showarticletitle{Design and Evaluation of GPU-FPX: A Low-Overhead Tool for Floating-Point Exception Detection in NVIDIA GPUs}. In \bibinfo{booktitle}{\emph{Proceedings of the 32nd International Symposium on High-Performance Parallel and Distributed Computing}} (Orlando, FL, USA) \emph{(\bibinfo{series}{HPDC '23})}. \bibinfo{publisher}{Association for Computing Machinery}, \bibinfo{address}{New York, NY, USA}, \bibinfo{pages}{59–71}.
\newblock
\showISBNx{9798400701559}
\urldef\tempurl%
\url{https://doi.org/10.1145/3588195.3592991}
\showDOI{\tempurl}


\bibitem[Loosemore et~al\mbox{.}(2019)]%
        {2019_loosemore}
\bibfield{author}{\bibinfo{person}{Sandra Loosemore}, \bibinfo{person}{Richard~M. Stallman}, {and} \bibinfo{person}{Andrew Oram}.} \bibinfo{year}{2019}\natexlab{}.
\newblock \bibinfo{title}{The GNU C Library Reference ManualThe GNU C Library Reference Manual}.
\newblock
\newblock


\bibitem[Menon et~al\mbox{.}(2018)]%
        {adapt}
\bibfield{author}{\bibinfo{person}{Harshitha Menon}, \bibinfo{person}{Michael~O. Lam}, \bibinfo{person}{Daniel Osei-Kuffuor}, \bibinfo{person}{Markus Schordan}, \bibinfo{person}{Scott Lloyd}, \bibinfo{person}{Kathryn Mohror}, {and} \bibinfo{person}{Jeffrey Hittinger}.} \bibinfo{year}{2018}\natexlab{}.
\newblock \showarticletitle{ADAPT: Algorithmic Differentiation Applied to Floating-Point Precision Tuning}. In \bibinfo{booktitle}{\emph{Proceedings of the International Conference for High Performance Computing, Networking, Storage, and Analysis}} (Dallas, Texas) \emph{(\bibinfo{series}{SC ’18})}. \bibinfo{publisher}{IEEE Press}, Article \bibinfo{articleno}{48}, \bibinfo{numpages}{13}~pages.
\newblock


\bibitem[Molecular Dynamics({[n.\,d.]})]%
        {MD}
Molecular Dynamics \bibinfo{year}{[n.\,d.]}\natexlab{}.
\newblock \bibinfo{title}{Molecular Dynamics}.
\newblock \bibinfo{howpublished}{\url{https://people.sc.fsu.edu/~jburkardt/py\_src/md/md.html}}.
\newblock


\bibitem[Moses et~al\mbox{.}(2021)]%
        {2021_enzyme}
\bibfield{author}{\bibinfo{person}{William~S. Moses}, \bibinfo{person}{Valentin Churavy}, \bibinfo{person}{Ludger Paehler}, \bibinfo{person}{Jan H\"{u}ckelheim}, \bibinfo{person}{Sri Hari~Krishna Narayanan}, \bibinfo{person}{Michel Schanen}, {and} \bibinfo{person}{Johannes Doerfert}.} \bibinfo{year}{2021}\natexlab{}.
\newblock \showarticletitle{Reverse-Mode Automatic Differentiation and Optimization of GPU Kernels via Enzyme}. In \bibinfo{booktitle}{\emph{Proceedings of the International Conference for High Performance Computing, Networking, Storage and Analysis}} (St. Louis, Missouri) \emph{(\bibinfo{series}{SC '21})}. \bibinfo{publisher}{Association for Computing Machinery}, \bibinfo{address}{New York, NY, USA}, Article \bibinfo{articleno}{61}, \bibinfo{numpages}{16}~pages.
\newblock
\showISBNx{9781450384421}
\urldef\tempurl%
\url{https://doi.org/10.1145/3458817.3476165}
\showDOI{\tempurl}


\bibitem[Muller et~al\mbox{.}(2010)]%
        {2010_Muller}
\bibfield{author}{\bibinfo{person}{Jean-Michel Muller}, \bibinfo{person}{Nicolas Brisebarre}, \bibinfo{person}{Florent Dinechin}, \bibinfo{person}{Claude-Pierre Jeannerod}, \bibinfo{person}{Vincent Lefèvre}, \bibinfo{person}{Guillaume Melquiond}, \bibinfo{person}{Nathalie Revol}, \bibinfo{person}{Damien Stehlé}, {and} \bibinfo{person}{Serge Torres}.} \bibinfo{year}{2010}\natexlab{}.
\newblock \bibinfo{booktitle}{\emph{Handbook of Floating-Point Arithmetic}}.
\newblock \bibinfo{publisher}{{B}irkhauser {B}oston}.
\newblock
\showISBNx{978-0-8176-4704-9}
\urldef\tempurl%
\url{https://doi.org/10.1007/978-0-8176-4705-6}
\showDOI{\tempurl}


\bibitem[Neumaier(2003)]%
        {taylorforms}
\bibfield{author}{\bibinfo{person}{Arnold Neumaier}.} \bibinfo{year}{2003}\natexlab{}.
\newblock \showarticletitle{Taylor Forms---Use and Limits}.
\newblock \bibinfo{journal}{\emph{Reliable Computing}} \bibinfo{volume}{9}, \bibinfo{number}{1} (\bibinfo{date}{Feb} \bibinfo{year}{2003}), \bibinfo{pages}{43--79}.
\newblock
\showISSN{1573-1340}
\urldef\tempurl%
\url{https://doi.org/10.1023/A:1023061927787}
\showDOI{\tempurl}


\bibitem[Numerics(2014)]%
        {darulova-phd}
Numerics \bibinfo{year}{2014}\natexlab{}.
\newblock \bibinfo{title}{Programming with Numerical Uncertainities}.
\newblock
\newblock
\urldef\tempurl%
\url{https://people.mpi-sws.org/~eva/papers/thesis.pdf}
\showURL{%
\tempurl}


\bibitem[Panchekha et~al\mbox{.}(2015)]%
        {herbie}
\bibfield{author}{\bibinfo{person}{Pavel Panchekha}, \bibinfo{person}{Alex Sanchez-Stern}, \bibinfo{person}{James~R. Wilcox}, {and} \bibinfo{person}{Zachary Tatlock}.} \bibinfo{year}{2015}\natexlab{}.
\newblock \showarticletitle{{Automatically Improving Accuracy for Floating Point Expressions}}. In \bibinfo{booktitle}{\emph{PLDI 2015}} \emph{(\bibinfo{series}{PLDI '15})}. \bibinfo{publisher}{ACM}.
\newblock
\newblock
\shownote{to appear \url{http://herbie.uwplse.org/}}.


\bibitem[Rubio-Gonz\'{a}lez et~al\mbox{.}(2013)]%
        {sc13-precimonious}
\bibfield{author}{\bibinfo{person}{Cindy Rubio-Gonz\'{a}lez}, \bibinfo{person}{Cuong Nguyen}, \bibinfo{person}{Hong~Diep Nguyen}, \bibinfo{person}{James Demmel}, \bibinfo{person}{William Kahan}, \bibinfo{person}{Koushik Sen}, \bibinfo{person}{David~H. Bailey}, \bibinfo{person}{Costin Iancu}, {and} \bibinfo{person}{David Hough}.} \bibinfo{year}{2013}\natexlab{}.
\newblock \showarticletitle{Precimonious: Tuning Assistant for Floating-Point Precision}. In \bibinfo{booktitle}{\emph{Supercomputing (SC)}}. \bibinfo{pages}{27:1--27:12}.
\newblock
\newblock
\shownote{\url{https://github.com/corvette-berkeley/precimonious}}.


\bibitem[Sagebaum et~al\mbox{.}(2017)]%
        {codipack}
\bibfield{author}{\bibinfo{person}{Max Sagebaum}, \bibinfo{person}{Tim Albring}, {and} \bibinfo{person}{Nicolas~R. Gauger}.} \bibinfo{year}{2017}\natexlab{}.
\newblock \showarticletitle{High-Performance Derivative Computations using CoDiPack}.
\newblock \bibinfo{journal}{\emph{CoRR}}  \bibinfo{volume}{abs/1709.07229} (\bibinfo{year}{2017}).
\newblock
\showeprint[arxiv]{1709.07229}
\urldef\tempurl%
\url{http://arxiv.org/abs/1709.07229}
\showURL{%
\tempurl}


\bibitem[Solovyev et~al\mbox{.}(2018)]%
        {fptaylor}
\bibfield{author}{\bibinfo{person}{Alexey Solovyev}, \bibinfo{person}{Marek~S. Baranowski}, \bibinfo{person}{Ian Briggs}, \bibinfo{person}{Charles Jacobsen}, \bibinfo{person}{Zvonimir Rakamariundefined}, {and} \bibinfo{person}{Ganesh Gopalakrishnan}.} \bibinfo{year}{2018}\natexlab{}.
\newblock \showarticletitle{Rigorous Estimation of Floating-Point Round-Off Errors with Symbolic Taylor Expansions}.
\newblock \bibinfo{journal}{\emph{ACM Trans. Program. Lang. Syst.}} \bibinfo{volume}{41}, \bibinfo{number}{1}, Article \bibinfo{articleno}{2} (\bibinfo{date}{Dec.} \bibinfo{year}{2018}), \bibinfo{numpages}{39}~pages.
\newblock
\showISSN{0164-0925}
\urldef\tempurl%
\url{https://doi.org/10.1145/3230733}
\showDOI{\tempurl}


\bibitem[symengine(2019)]%
        {symengine}
symengine \bibinfo{year}{2019}\natexlab{}.
\newblock \bibinfo{title}{SymEngine}.
\newblock \bibinfo{howpublished}{\url{https://github.com/symengine/symengine/}}.
\newblock


\bibitem[Tirpankar et~al\mbox{.}(2025)]%
        {2025_fpguard}
\bibfield{author}{\bibinfo{person}{Tanmay Tirpankar}, \bibinfo{person}{Artem Yadrov}, {and} \bibinfo{person}{Ganesh Gopalakrishnan}.} \bibinfo{year}{2025}\natexlab{}.
\newblock \bibinfo{title}{FPGUARD: Static-Analysis Guided Domain Exclusions for Robust Floating-Point Error Analysis (to appear)}.
\newblock
\newblock


\bibitem[Titolo et~al\mbox{.}(2017)]%
        {precisa}
\bibfield{author}{\bibinfo{person}{Laura Titolo}, \bibinfo{person}{Marco~A. Feli{\'{u}}}, \bibinfo{person}{Mariano Moscato}, {and} \bibinfo{person}{C{\'{e}}sar~A. Mu{\~{n}}oz}.} \bibinfo{year}{2017}\natexlab{}.
\newblock \showarticletitle{An Abstract Interpretation Framework for the Round-Off Error Analysis of Floating-Point Programs}.
\newblock In \bibinfo{booktitle}{\emph{Lecture Notes in Computer Science}}. \bibinfo{publisher}{Springer International Publishing}, \bibinfo{pages}{516--537}.
\newblock
\urldef\tempurl%
\url{https://doi.org/10.1007/978-3-319-73721-8_24}
\showDOI{\tempurl}


\bibitem[Zuras(2008)]%
        {2008_zuras}
\bibfield{author}{\bibinfo{person}{Dan Zuras}.} \bibinfo{year}{2008}\natexlab{}.
\newblock \showarticletitle{IEEE Standard for Floating-Point Arithmetic}.
\newblock \bibinfo{journal}{\emph{IEEE Std 754-2008}} \bibinfo{number}{N/A} (\bibinfo{year}{2008}), \bibinfo{pages}{1--70}.
\newblock
\urldef\tempurl%
\url{https://doi.org/10.1109/IEEESTD.2008.4610935}
\showDOI{\tempurl}


\end{thebibliography}

\end{document}